\documentclass[10pt, notitlepage]{revtex4-1} 

\usepackage{tikz}
\usepackage{amsmath}
\usepackage{mathtools}
\usepackage{amssymb}
\usepackage{lmodern}
\usepackage{geometry}
\usepackage{graphicx}
\usepackage{verbatim}
\usepackage{epstopdf}
\usepackage{array}
\usepackage{epsfig}

\newcommand\Tstrut{\rule{0pt}{2.6ex}}
\newcommand*\colvec[3][]{
    \left[ \; \begin{matrix}\ifx\relax#1\relax\else#1\\[3pt]\fi#2\\[3pt]#3\end{matrix} \;  \right]
}

\newcommand*\colvecr[3][]{
    \left[ \; \begin{matrix*}[r]\ifx\relax#1\relax\else#1\\[3pt]\fi#2\\[3pt]#3\end{matrix*} \;  \right]
}

\begin{document}
\title{Statistical Topology of Perturbed Two-Dimensional Lattices}
\author{Hannes Leipold$^1$, Emanuel A. Lazar$^1$, Kenneth A. Brakke$^2$, David J. Srolovitz$^{1,3}$}
\affiliation
{$^1$Department of Materials Science and Engineering, \\University of Pennsylvania, Philadelphia, PA 19104 \\
$^2$Department of Mathematical Sciences, \\Susquehanna University, Selinsgrove, PA 17870\\
$^3$Department of Mechanical Engineering and Applied Mechanics,\\ University of Pennsylvania, Philadelphia, PA 19104 
}
\date{\today}

\begin{abstract} 
The Voronoi cell of any atom in a lattice is identical.  If atoms are perturbed from their lattice coordinates, then the topologies of the Voronoi cells of the atoms will change.  We consider the distribution of Voronoi cell topologies in two-dimensional perturbed systems.  These systems can be thought of as simple models of finite-temperature crystals. We give analytical results for the distribution of Voronoi topologies of points in two-dimensional Bravais lattices under infinitesimal perturbations and present a discussion with numerical results for finite perturbations. 
\end{abstract}

\maketitle

\section{Introduction}
\subsection{Motivation}

Statistical topology provides a set of tools for studying the statistics of topological properties of a system, much in the way that statistical mechanics focuses on geometric quantities such as velocities and momenta of large sets of particles.  Statistical topology has been applied to many classical systems including glasses \cite{rivier1983statistical}, polymers \cite{orlandini2007statistical}, radial and cellular networks \cite{seong2012statistical, mason2012statistical}, and ideal gases \cite{2013lazar}.  In this paper we consider the statistical topology of perturbed lattices.  In particular, we consider systems in which atoms are initially located at lattice positions and then perturbed by random displacements; such perturbations might be associated with thermal vibrations of the atoms.  In two dimensions, the complete topology of each Voronoi cell is given by its number of sides $n$.  What is the distribution of $n$?  We provide exact results for each of the Bravais lattices and partial results for the honeycomb structure.  Our goal is to provide a more complete understanding of realistic systems by providing an analytic point of comparison.

{\bf Statement of Problem.}  We begin by considering a system of atoms arranged on a two-dimensional lattice
\begin{equation}
\Lambda = \left\{ n_1 {\bf v_1} + n_2 {\bf v_2} \;  | \; n_1, n_2 \in \mathbb{Z} \right \}, 
\end{equation}
where $\bf v_1$ and $\bf v_2$ are linearly independent. A random perturbation of this lattice is obtained by displacing each atom from its lattice position by a random variable.  Properties of such systems are considered in \cite{holroyd2013insertion, peres2014rigidity}.  This system can be considered as an approximation for a two-dimensional crystalline solid whose atoms are displaced from lattice positions by thermal noise and small strains.  For a given lattice $\Lambda$, we consider $p(n)$, the probability that a randomly chosen atom will have a Voronoi cell with $n$ edges.  

In this paper, we consider perturbations which are chosen from a radially symmetric distribution, such that the probability of a particular displacement from a lattice position depends only on its magnitude.  One particular example of such a model is the classical Einstein solid, in which atoms are perturbed from lattice positions using a Gaussian distribution \cite{kittel1976introduction}. The Gaussian distribution arises from atomic vibrations within a parabolic potential well, where the width of the distribution is proportional to the square root of the temperature. We begin by obtaining analytic results for infinitesimal perturbations and later consider perturbations of finite amplitude. 
\begin{figure}
\begin{center}$
\begin{tabular}{ccc}
\fbox{\includegraphics[height=0.25\linewidth]{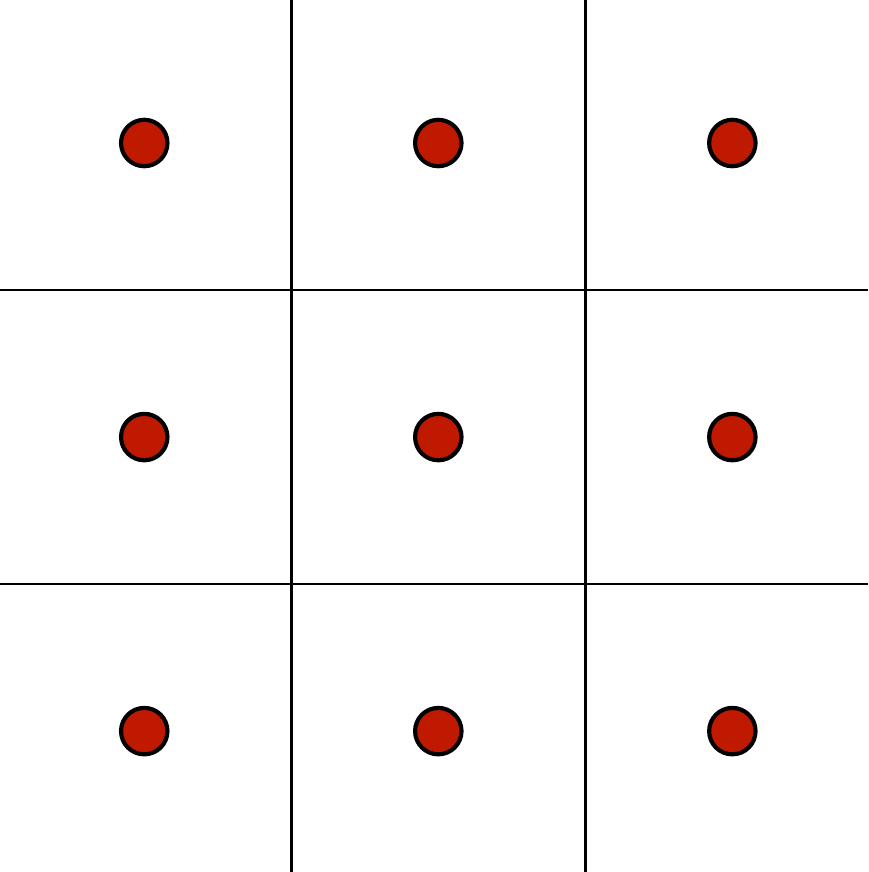}} &
\fbox{\includegraphics[height=0.25\linewidth]{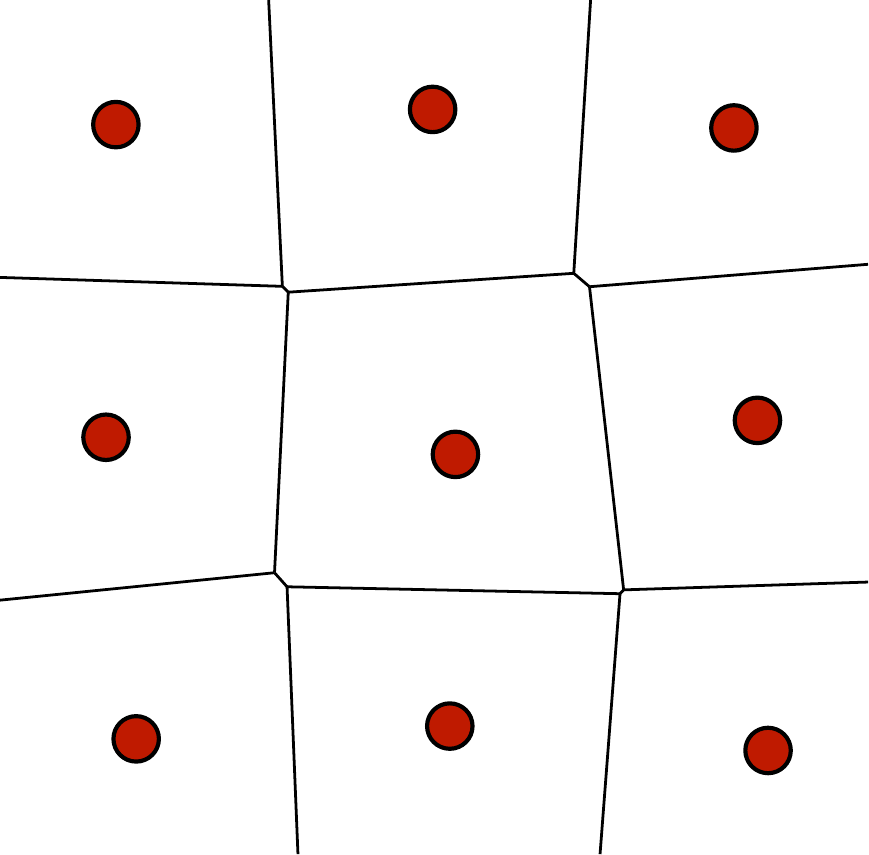}} &
\fbox{\includegraphics[height=0.25\linewidth]{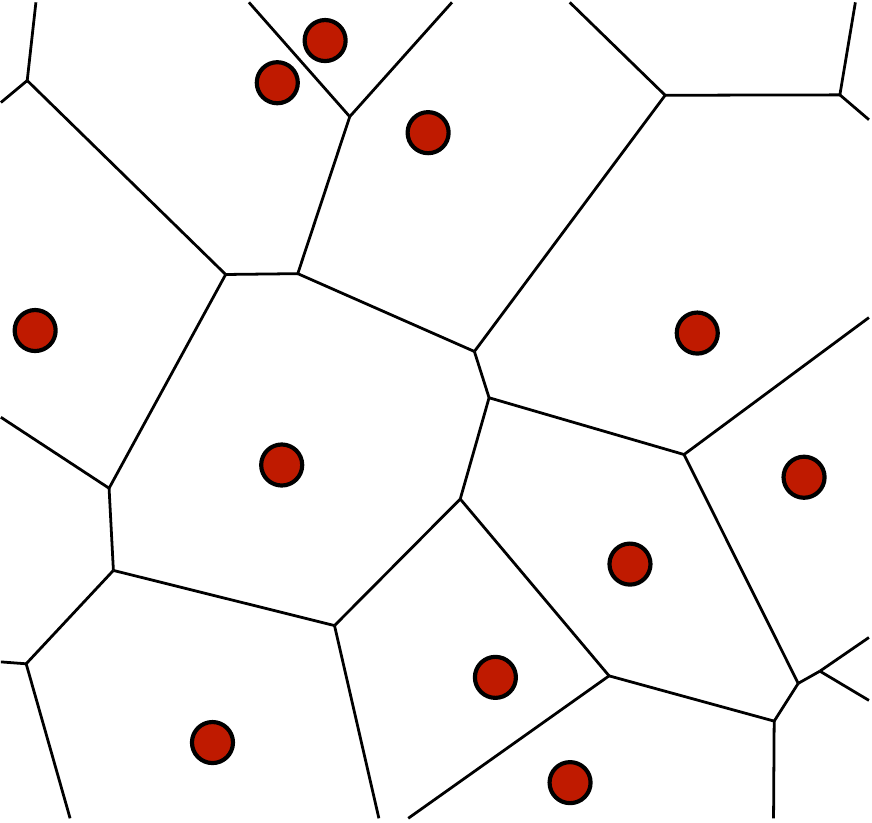}} \vspace{1mm}\\
(a) & (b) & (c)
\end{tabular}$
\end{center}
\caption{Sets of atoms and their Voronoi tessellations: (a) square lattice, (b) perturbed square lattice, and (c) random Poisson point process.}
\label{examples}
\end{figure}

\subsection{Voronoi Tessellations}
\label{Voronoi Tessellations}

We consider Voronoi tessellations in $\mathbb{R}^2$ endowed with the standard Euclidean metric $d$.  Let $S$ be a discrete, possibly infinite, set of atomic coordinates in $\mathbb{R}^2$.  Then,
\begin{equation}
R_s = \{x \in \mathbb{R}^2 \;|\; d(x, s) \leq d(x, s')\; \text{for all}\; s' \in S\}
\end{equation}
is the Voronoi cell associated with an atom $s\in S$.  In other words, the Voronoi tessellation is the partitioning of the plane into regions such that all points in a region are not closer to any other atom than to its own.  Note that a point $x$ can be equidistant to multiple atoms and hence belong to multiple Voronoi cells.  Figure \ref{examples} shows Voronoi tessellations for a square lattice, a perturbed square lattice, and a set of Poisson-distributed atoms.  For our purposes, atoms are said to be in general position if no more than three atoms lie on the boundary of an {\it empty} circle, i.e., a circle that has no other atoms within it.  If atoms are in general position, then no point belongs to more than three distinct cells, and so no point is adjacent to more than three edges; note that in the square lattice this is not the case.  In Figs.~\ref{examples}(b) and (c), exactly three edges meet at any point.  In this case, Euler's formula for planar graphs \cite{west2001introduction} requires that the average number of edges per cell be exactly six.  

\begin{figure}
\begin{center}
\begin{tabular}{ccc}
\fbox{\includegraphics[height=0.22\linewidth]{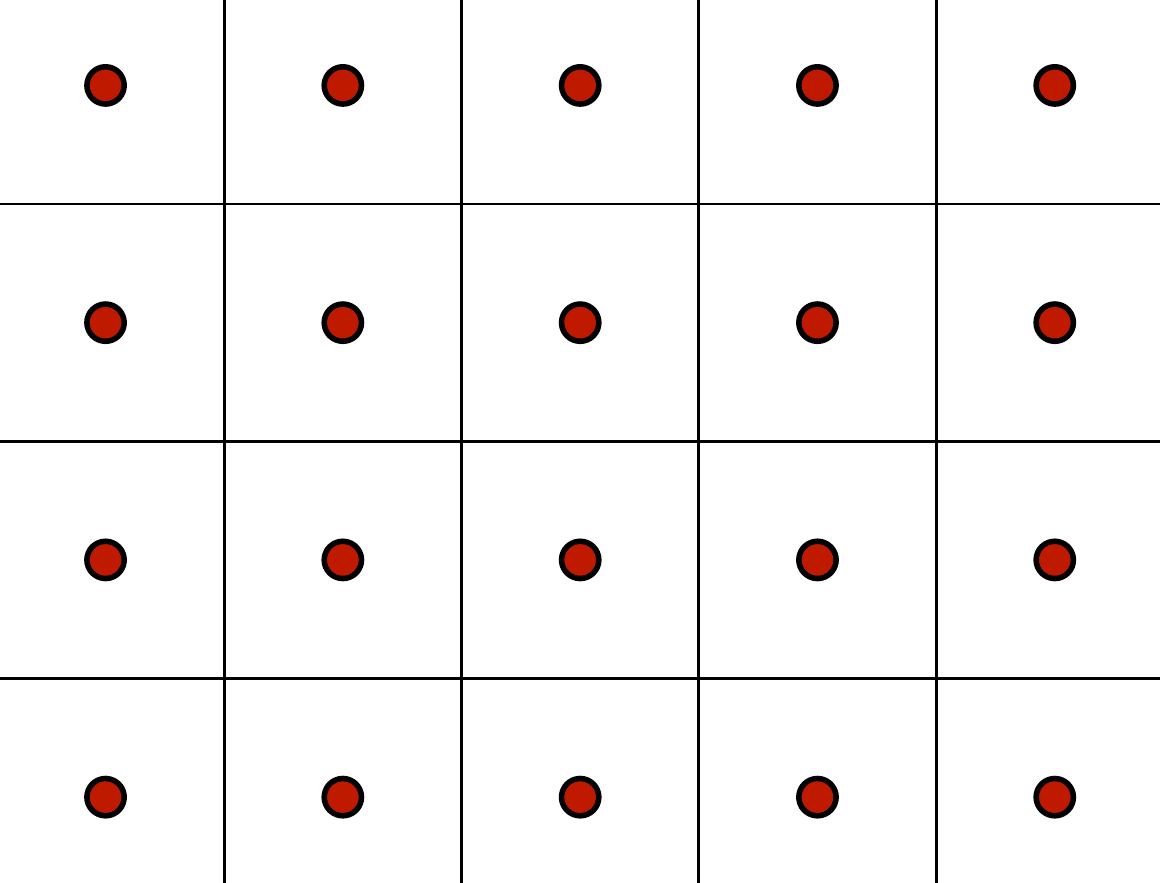}} &
\fbox{\includegraphics[height=0.22\linewidth]{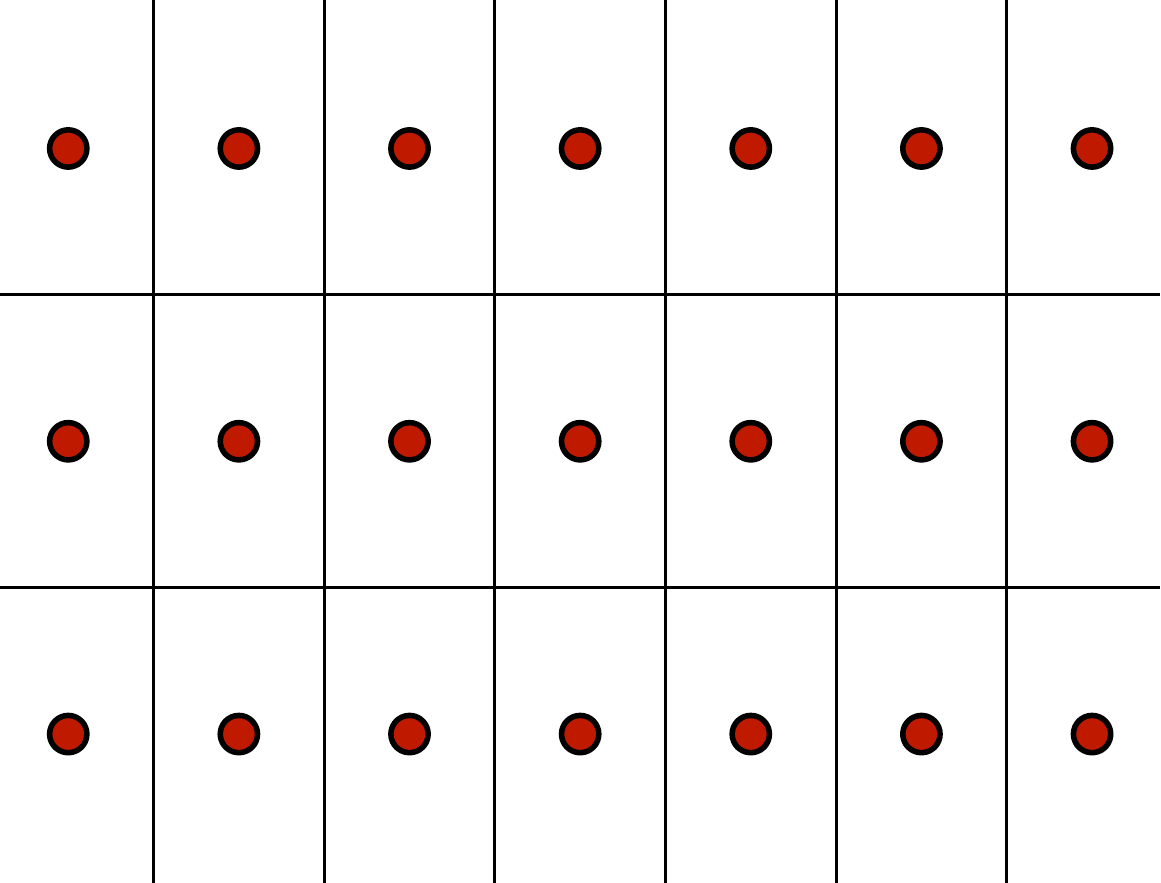}} &
\fbox{\includegraphics[height=0.22\linewidth]{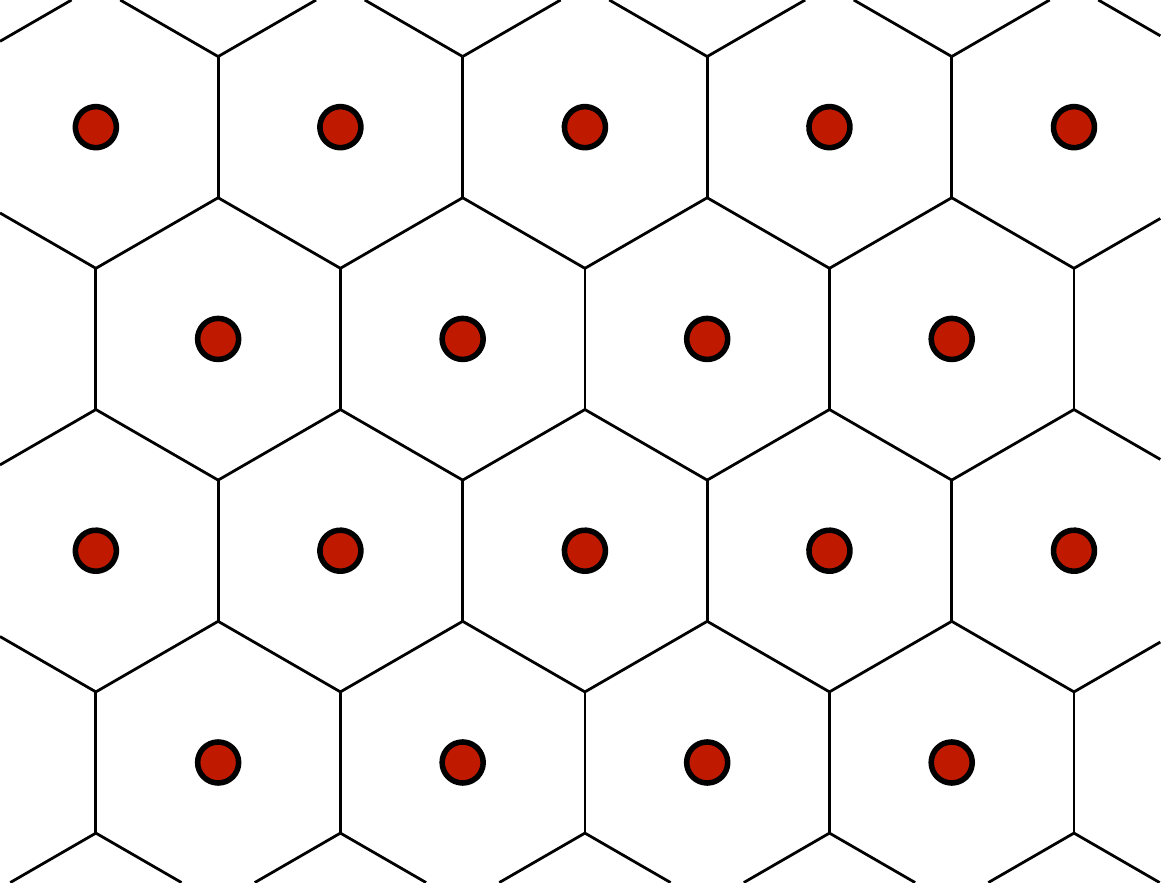}} \vspace{1mm}\\
(a) Square & (b) Rectangular & (c) Hexagonal \vspace{2mm}\\
\fbox{\includegraphics[height=0.22\linewidth]{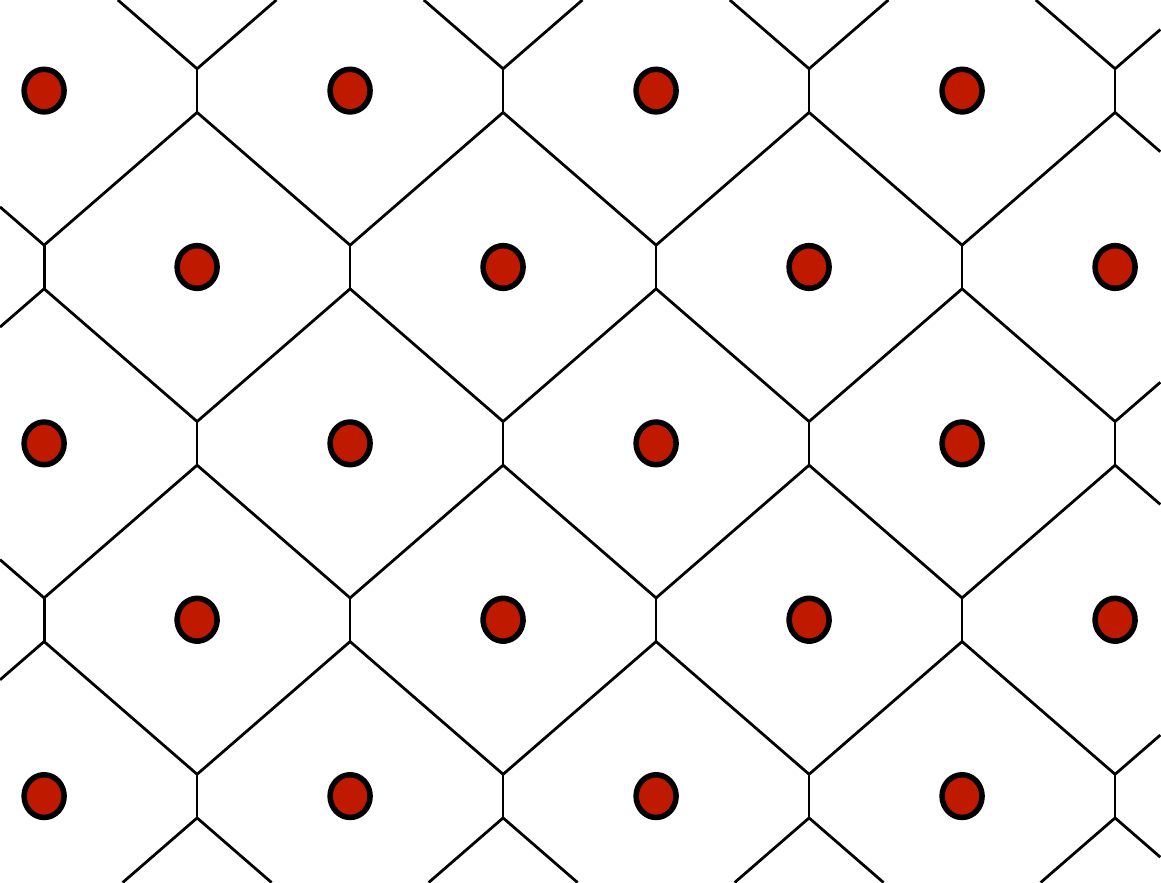}} &
\fbox{\includegraphics[height=0.22\linewidth]{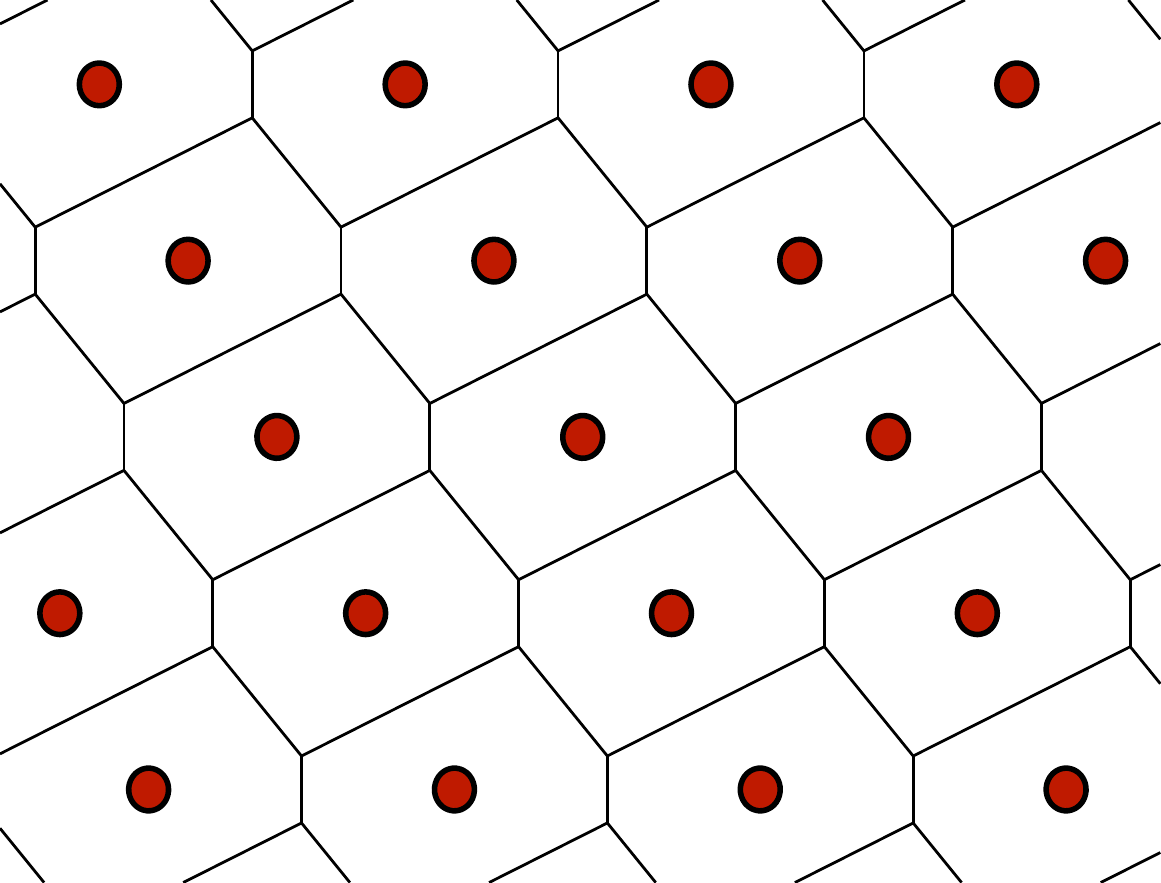}} &
\fbox{\includegraphics[height=0.22\linewidth]{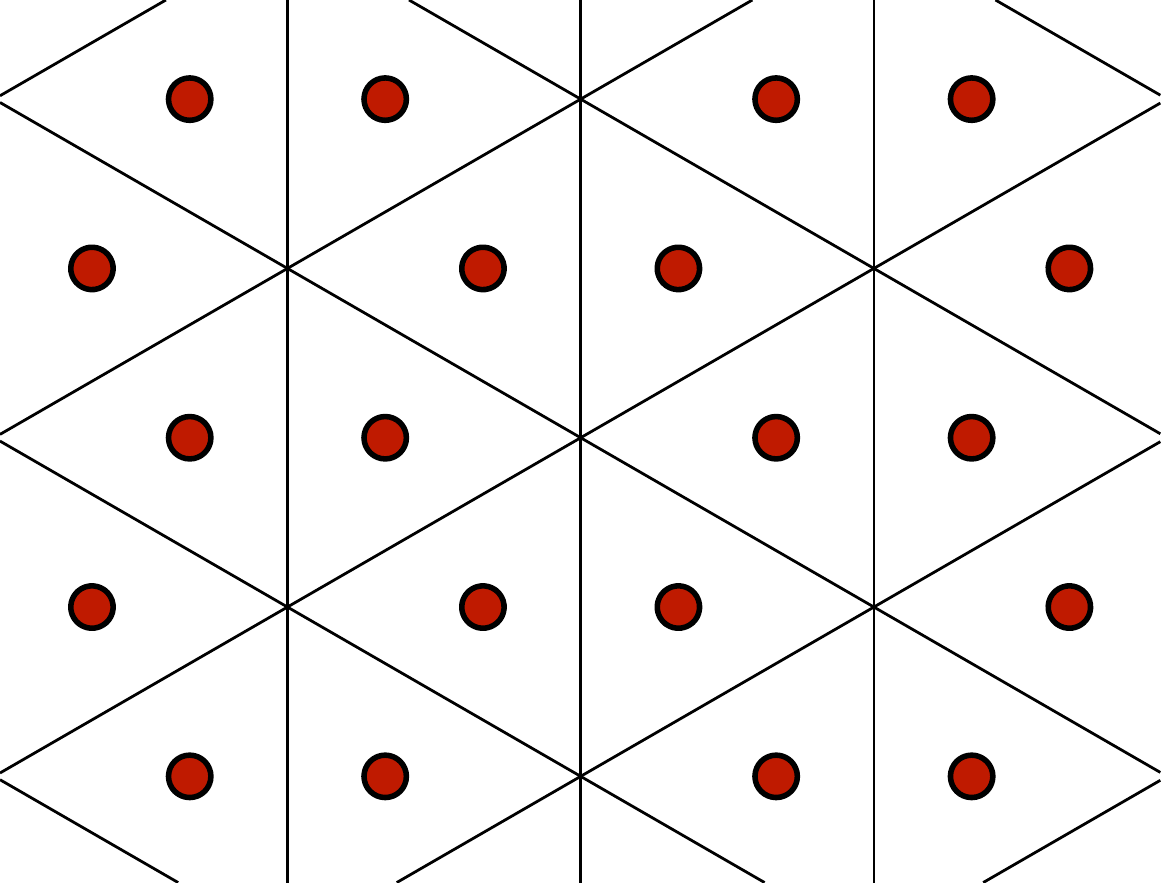}} \vspace{1mm}\\
(d) Rhombic & (e) Oblique & (f) Honeycomb (non-lattice) \\
\end{tabular}
\end{center}
\caption{(a-e) All five two-dimensional Bravais lattices and their Voronoi tessellations. (f) The honeycomb structure (not a Bravais lattice, since there are atoms not equivalent by translation) and its Voronoi tessellation.\label{bravais}}
\end{figure}

\subsection{Classification of Two Dimensional Lattices}
\label{Classification for Two Dimensional Lattices}

Two-dimensional lattices are traditionally classified into five groups, known as the Bravias lattices \cite{bravais1866etudes} and illustrated in Fig.~\ref{bravais}.  This classification is based on the relative magnitudes and orientations of ${\bf v_1}$ and ${\bf v_2}$.  If ${\bf v_1}$ and ${\bf v_2}$ are orthogonal, then the lattice is either {\it square} if $\|{\bf v_1}\| = \|{\bf v_2}\|$ or {\it rectangular} otherwise.  If the angle between the vectors is $60^\circ$ or $120^\circ$ and $\|{\bf v_1}\| = \|{\bf v_2}\|$, then the lattice is {\it hexagonal}.  For all other angles, the lattice is either {\it rhombic} or {\it oblique}, depending on whether $(2{\bf v_2}-{\bf v_1})\cdot {\bf v_1}=0$.

Each lattice has a Voronoi cell which shares the symmetry group of the lattice itself and which can be repeated to tile the plane.  The square and rectangular lattices have four-sided cells, while the hexagonal, rhombic, and oblique lattices each have six-sided cells.  The Voronoi tessellation of square and rectangular lattices are {\it unstable} in the sense that infinitesimal perturbations of the atom positions will change the topology of the Voronoi structure.  That is because in these lattices, vertices in the Voronoi tessellation are equidistant to four atoms.  In contrast, the hexagonal, rhombic and oblique tessellations are stable, and no more than three edges meet at any vertex.

\subsection{Delaunay Triangulation}
\label{Delaunay Triangulation}

\begin{figure}
\setlength{\fboxsep}{0pt}
\begin{center}
\fbox{\includegraphics[height=0.25\linewidth]{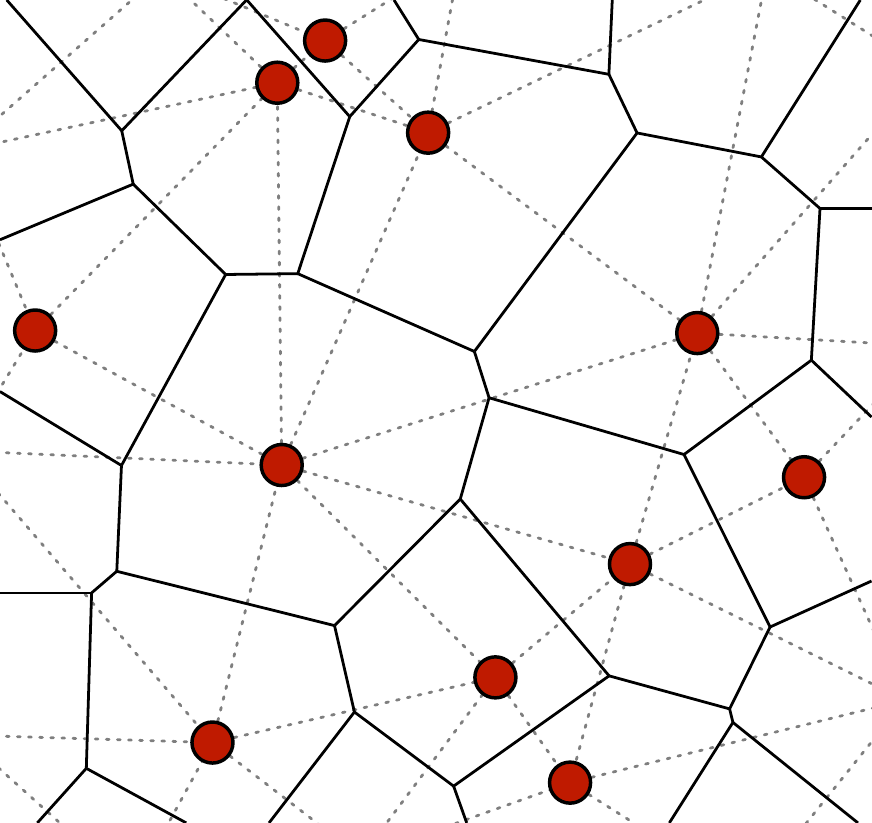}} 
\end{center}
\caption{A set of atoms, the associated Voronoi tessellation (solid lines), and the Delaunay triangulation (dashed lines); each Voronoi edge is perpendicular to a Delaunay edge.}
\label{voro-delaunay}
\end{figure}

Every Voronoi tessellation has a dual object called a Delaunay diagram.  In a two-dimensional Delaunay diagram, edges connect atoms whose Voronoi cells share an edge.  If atoms are in general position (i.e., no four lie on the boundary of an empty circle), then the Delaunay diagram will be a triangulation.  Figure \ref{voro-delaunay} illustrates a set of atoms, the Voronoi tessellation of the set, and its dual Delaunay diagram.

The Delaunay diagram is a useful tool for computing properties of its dual Voronoi tessellation.  A Delaunay triangulation maximizes the minimum interior angle of each triangle \cite{de2000computational}.  Simple mechanisms for finding the Delaunay triangulation, and thus the Voronoi tessellation of a set of atoms in general position, are the flip algorithms considered in \cite{de2000computational}. These algorithms begin by constructing a triangulation in an arbitrary manner.  Next, for each pair of adjacent triangles, the edge separating the pair is flipped and it is checked whether this increases the minimal internal angle in those triangles.  If this is indeed the case, then the edge flip is retained; otherwise it is reversed.  This ensures that the adjacent triangles are {\it locally} Delaunay. This process is repeated until no pair of adjacent triangles is non-Delaunay.   
\begin{figure}
\begin{center}
\includegraphics[height=0.27\linewidth]{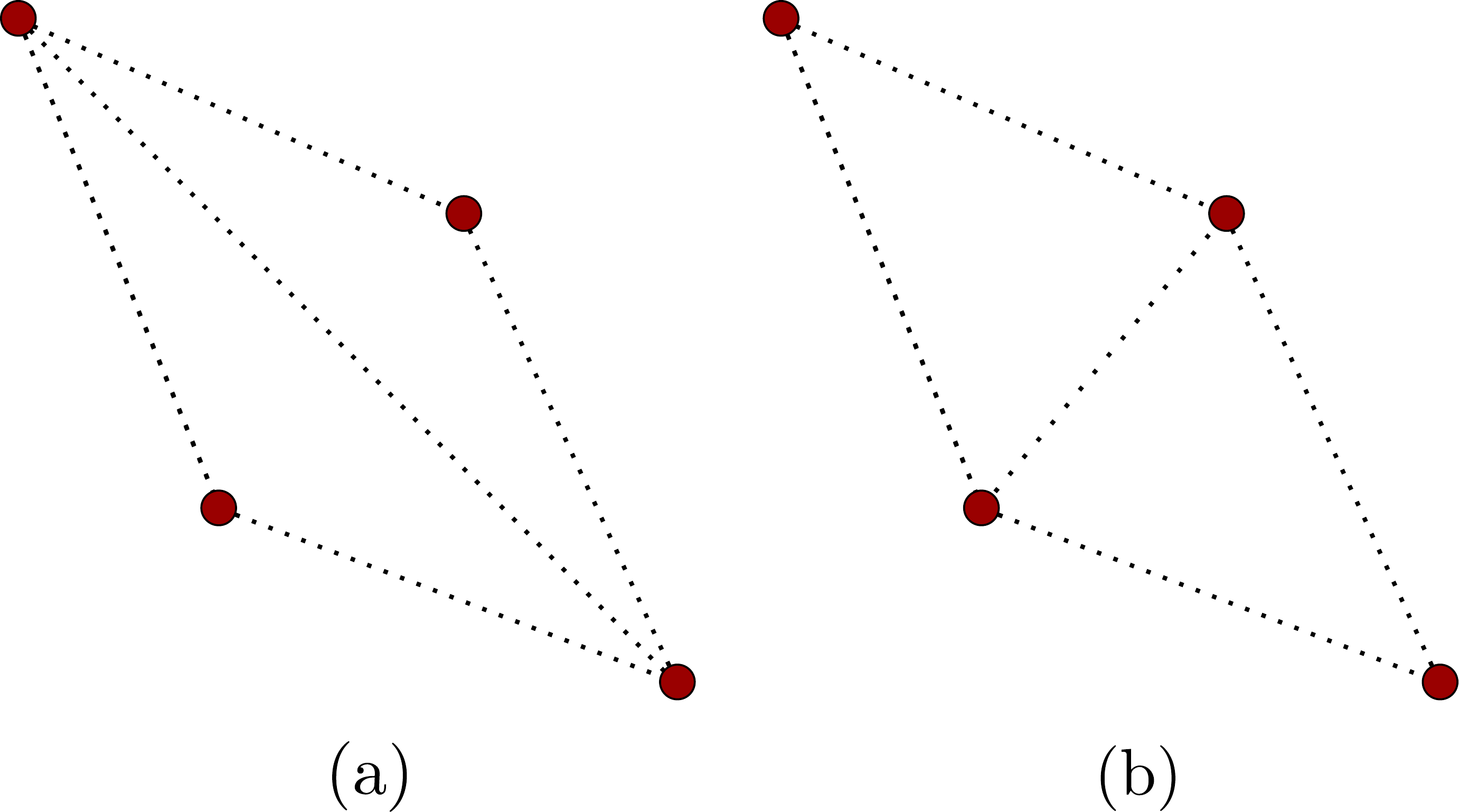}
\vspace*{-\baselineskip}
\end{center}
\caption{(a) A local non-Delaunay triangulation; (b) the corresponding local Delaunay triangulation.\label{triangulations}}
\vspace*{-\baselineskip}
\end{figure}

Given the two triangulations for a region bounded by four atoms, picking the one that maximizes the minimum internal angle ensures the Delaunay condition is met locally. A simple corollary to this theorem is that if a pair of opposite angles adds to greater than $ 180^{\circ} $ on a quadrilateral, then the Delaunay triangles split this pair. In Fig. \ref{triangulations}, it is clear that (a) has a smaller minimum angle than does (b); its is also clear that two opposite angles add to greater than $180^{\circ} $ in (a) but not in (b).  Therefore, Fig. \ref{triangulations}(b) is Delaunay while \ref{triangulations}(a) is not.

\section{Analytic Solutions}
\label{Analytic Solutions}

\subsection{Vertex in a Square Lattice}
\label{Vertex in a Square Lattice}

\begin{figure}
\includegraphics[width=0.68\linewidth]{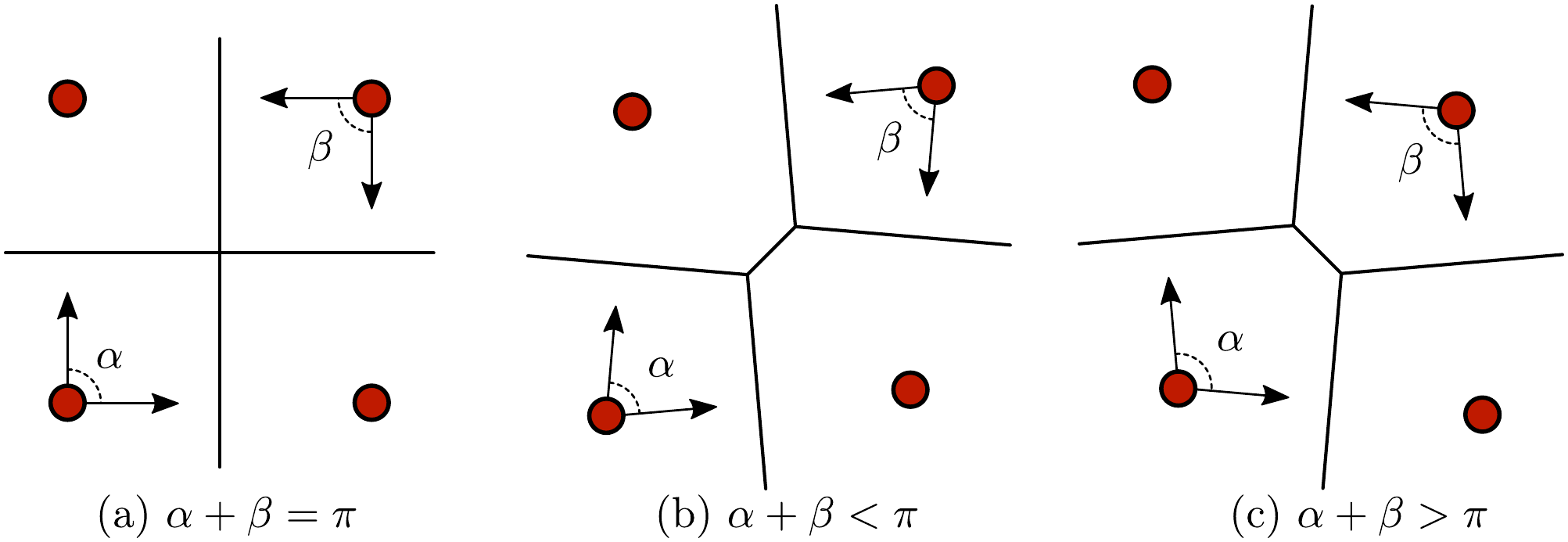}
\caption{Four atoms and the nearby Voronoi tessellation (shown as solid lines).  (a) Before perturbations, all internal angles of the Delaunay diagram (Voronoi dual), including $\alpha=\beta$, are equal to $\pi/2$. After perturbations, the Voronoi cell and Delaunay diagram change such that  either (b) $\alpha+\beta<\pi$ and a Voronoi edge forms in one direction, or else (c) $\alpha+\beta>\pi$ and a Voronoi edge forms in the other direction.\label{pert_square}}
\end{figure}

Before calculating the distribution of Voronoi topologies $p(n)$ for perturbed lattices, we consider how the Voronoi tessellation near a topologically unstable vertex (i.e., a vertex with more than 3 edges) resolves under small perturbations of the atomic coordinates.  We focus first on the Voronoi tessellation of the square lattice illustrated in Fig.~\ref{pert_square}(a).  In the unperturbed case, some points in the plane are equidistant to four neighboring atoms and therefore lay on the boundaries of four Voronoi cells.  After atom positions are randomly perturbed, each unstable Voronoi vertex will resolve into two topologically-stable vertices.  Such a resolution can happen in two distinct ways, as illustrated in Figs.~\ref{pert_square}(b) and (c).  Given the symmetry of the problem, it seems intuitive that each resolution  should occur with equal probability.  Here we make this intuition precise.

As noted above, the Delaunay diagram can be used to study properties of its dual, the Voronoi tessellation.  In the unperturbed case illustrated in Fig.~\ref{pert_square}(a), all internal angles of the Delaunay diagram are $\pi/2$, including the pair $ \alpha, \beta$.  After a random perturbation, $\alpha + \beta$ will almost surely either decrease or increase, determining whether the unstable vertex resolves in the manner of Fig.~\ref{pert_square}(b) or \ref{pert_square}(c), respectively.  Determining the probability of obtaining these configurations thus requires calculating the probability with which $\alpha + \beta$ increases or decreases under a random perturbation.

In the remainder of the paper we denote a spatial configuration of $n$ atoms in $\mathbb{R}^2$ as
\begin{equation}
X = \left( \colvec{ x_1 }{ y_1 }, \colvec{ x_2 }{ y_2 }, \cdots , \colvec{ x_n }{ y_n } \right) \in \mathbb{R}^{2n}, 
\end{equation} 
such that the initial, unperturbed configuration of four atoms on a square lattice is 
\begin{equation}
X = \left( \colvec{ 0 }{ 0 }, \colvec{ 1 }{ 0 }, \colvec{ 1 }{ 1 }, \colvec{ 0 }{ 1 } \right).  \hphantom{\in \mathbb{R}^{2n}.}
\end{equation}
We use $\phi(X)$ to denote the sum of the angles $\alpha$ and $\beta$ illustrated in Fig.~\ref{pert_square}(a).  We apply a random perturbation to  $\phi$, namely $ \xi \in \mathbb{R}^{2n} $, where $ \xi $ is chosen from any radially symmetric probability distribution. As $\phi$ changes smoothly with  atomic coordinates, the small perturbation $\xi$ can be linearized  as $\nabla \phi \cdot \xi$, such that $ \phi(X + \xi) \approx \phi(X) + \nabla\phi \cdot \xi$.

In general, the set of perturbations $\xi$  that satisfy $\nabla \phi \cdot \xi=0$ can be represented as a hyperplane in $\mathbb{R}^{2n}$ passing through the origin. For perturbations $\xi$ on one side of this hyperplane we have $\nabla \phi \cdot \xi> 0$, while for those on the other side we have $\nabla \phi \cdot \xi< 0$. For the configuration to resolve as shown in Fig.~\ref{pert_square}(b), it must be that $ \phi(X+\xi) > \pi $ or, equivalently, $ \nabla \phi \cdot \xi > 0 $.  If $ \nabla \phi \cdot \xi < 0 $, the initial configuration will resolve as shown in Fig.~\ref{pert_square}(c). Since the distribution of perturbations is symmetric about the origin, the integral of the probability measure over the region for which $\nabla \phi \cdot \xi>0$ is 0.5; this is also the integral over the region for which $\nabla \phi \cdot \xi<0$.  This demonstrates that the topological resolutions illustrated in Figs.~\ref{pert_square}(b) and (c) indeed occur with equal probability.

In this example, the probabilities depend on the fractional volume of space on either side of a single hyperplane.  We next consider probabilities of more complex configurations, the calculations of which require consideration of fractional volumes associated with multiple intersecting half-spaces.

\subsection{Nearby Vertices in a Square Lattice}
\label{Two Nearby Vertices in a Square Lattice}

\begin{figure}
\begin{center}
\includegraphics[width=0.9\linewidth]{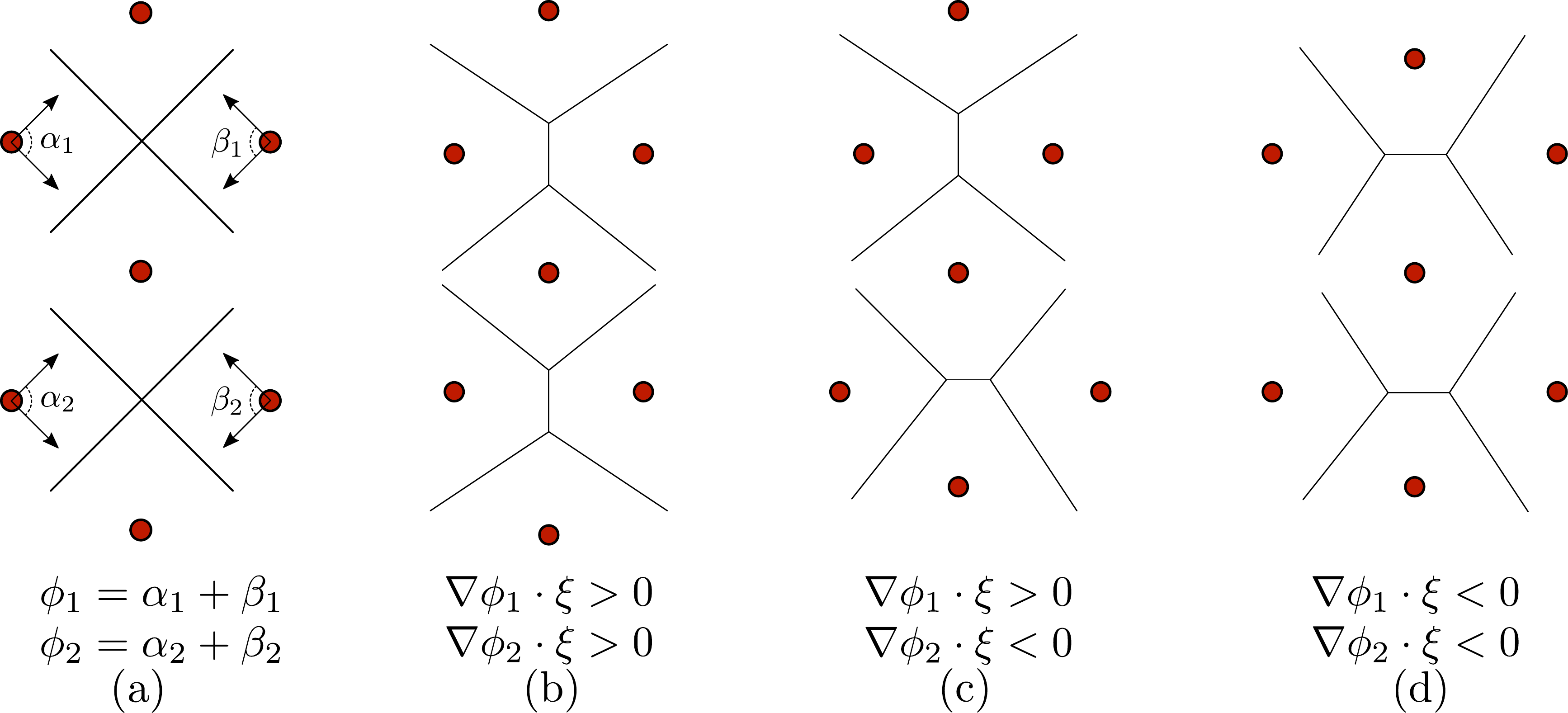}
\end{center}
\vspace*{-\baselineskip}
\caption{(a) Prior to perturbation, all internal angles of the Delaunay diagram are $\pi/2$, and the two vertices of the  Voronoi tessellation are unstable.  After perturbation, each vertex will resolve in a manner that depends on $\alpha_i+\beta_i$.  (b-d) show how the topology of the Delaunay diagram resolves depending on the signs of $\nabla\phi_1\cdot\xi$ and $\nabla\phi_2\cdot\xi$. Flipping the signs in (c) leads to the remaining resolution, which is identical to (c) up to rotation by $180^{\circ}$.\label{pert_square23}}
\end{figure}

Another example illustrates the complexities that can arise from correlations between neighboring unstable vertices.  Figure \ref{pert_square23} illustrates two nearby, unstable vertices in a Voronoi tessellation of a square lattice.  The manner in which these resolve is determined by the motions of seven nearby atoms, and how the labeled angles $\alpha_i$ and $\beta_i$ change under  perturbations.

\begin{figure}
\begin{center}
\includegraphics[width=0.6\linewidth]{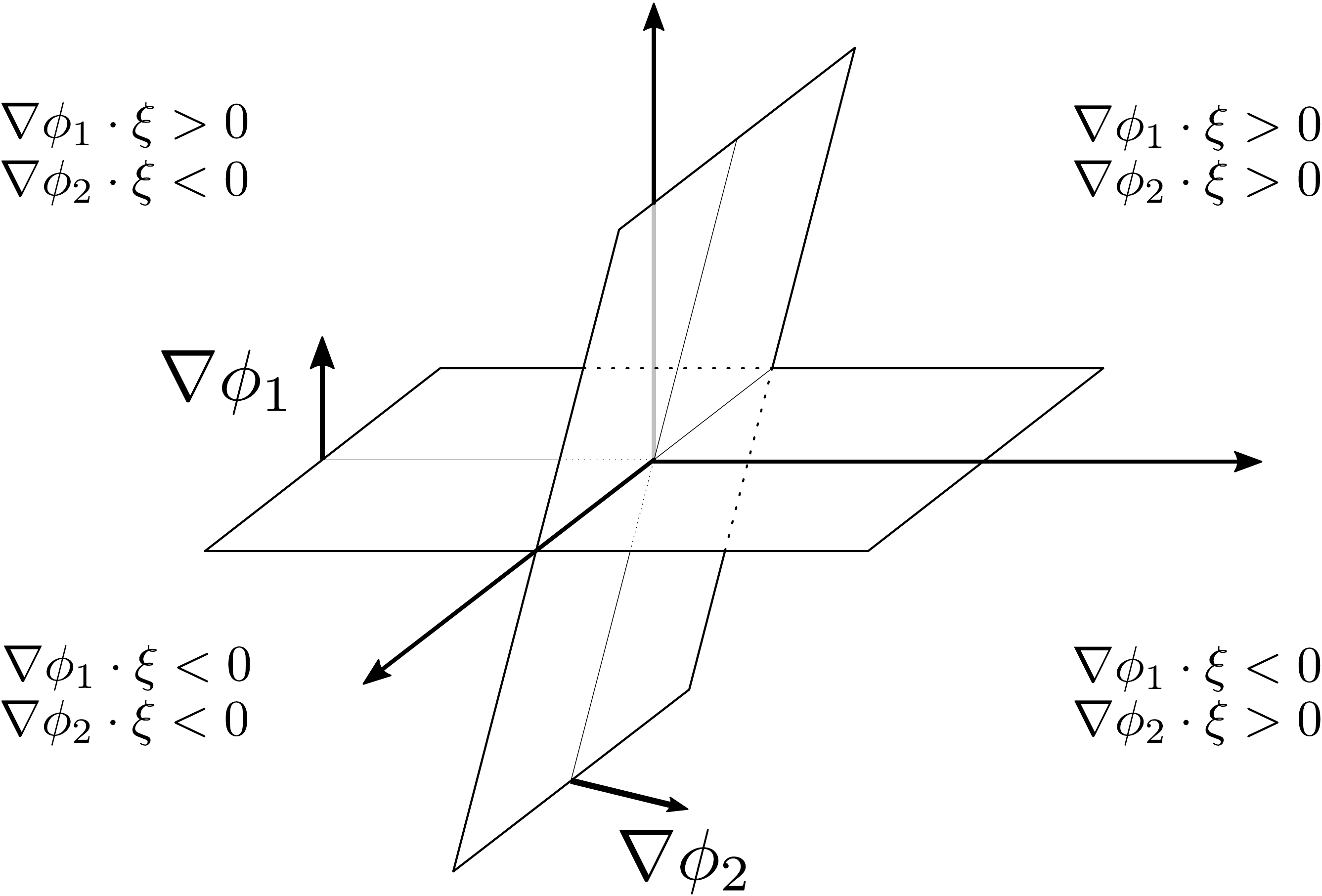}
\end{center}
\caption{Schematic of $\mathbb{R}^{2n}$ divided into regions in which the topology resolves in distinct manners.  Each $\nabla\phi_i$ divides $\mathbb{R}^{2n}$ into two regions, one on which $\nabla\phi_i\cdot\xi<0$, and the other on which $\nabla\phi_i\cdot\xi>0$, where $\xi$ is a point in $\mathbb{R}^{2n}$.\label{twoplanes}}
\end{figure}

Before perturbation $\alpha_i=\beta_i=\pi/2$ for all $i$, and the two vertices of the Voronoi tessellation are unstable.  After perturbation, if $\alpha_i+\beta_i>\pi$  a new edge will form between the two atoms, as illustrated in Fig.~\ref{pert_square}(c).  As shown above, the probability of $\alpha_i+\beta_i>\pi$ is 0.5.  What is the joint probability that both $\alpha_1+\beta_1$ and $\alpha_2+\beta_2$ will increase under a random perturbation?

As before, we denote a configuration of 7 atoms as a vector $X \in \mathbb{R}^{2n}$, where $n=7$; we let $\phi_i(X) = \alpha_i + \beta_i$.  Each of $\nabla\phi_i \cdot \xi =0$ determines a hyperplane through the origin in $\mathbb{R}^{2n}$, and the two intersecting hyperplanes divide $\mathbb{R}^{2n}$ into four regions, each corresponding to one of four possible topologies that result from the infinitesimal perturbations of the  atoms, as illustrated in Fig.~\ref{pert_square23}(b). If we let the initial, unperturbed configuration space of atoms be:
\begin{equation}
X = \left( \colvec{ 0 }{ 2 }, \colvecr{  -1 }{ 1 }, \colvec{  1 }{ 1 }, \colvec{  0 }{ 0 }, \colvec{ -1 }{ -1 }, \colvecr{ -1 }{ 1 }, \colvecr{ 0 }{ -2 } \right) 
\end{equation}
then the gradients $\nabla\phi_i$ are:
\begin{eqnarray}
\nabla \phi_1 &=& \left( \colvec{ 0 }{ 1 } , \colvec{ 1 }{ 0 } , \colvec{ 1 }{ 0 } , \colvecr{ 0 }{ -1 } , \colvec{ 0 }{ 0 } , \colvec{ 0 }{ 0 } , \colvec{ 0 }{ 0 } \right) {\text{   and}}\\ 
\nabla \phi_2 &=& \left( \colvec{ 0 }{ 0 } , \colvec{ 0 }{ 0 } , \colvec{ 0 }{ 0 } , \colvec{ 0 }{ 1 } , \colvec{ 1 }{ 0 } , \colvecr{ -1 }{ 0 } , \colvecr{ 0 }{ -1 } \right).
\end{eqnarray}
The two hyperplanes defined by these gradient vectors intersect at an angle $\theta = \arccos(-1/4) \approx 104.48^{\circ}$; this angle does not depend on the choice of coordinates.

Since the perturbations are chosen from a probability distribution that is radially symmetric about the origin, the probabilities of the configuration resolving into each of the four possibilities are equal to the fractional volumes of the four regions separated by the intersecting planes (see Fig.~\ref{twoplanes}).  These fractional volumes can be determined through consideration of the angles at which these two planes intersect.  

In the case considered here, the angle between $\nabla \phi_1$ and $\nabla \phi_2$ is $\theta = \arccos(-1/4)$, and so the angle subtended by the region where $\nabla \phi_1 \cdot \xi >0$ and $\nabla \phi_2 \cdot \xi >0$ is $ \pi - \theta $. The fractional volume of this region is then $ (\pi - \theta)/2\pi \approx 20.98\% $. The fractional volume of the region where $ \nabla \phi_1 \cdot \xi <0$ and $\nabla \phi_2 \cdot \xi <0$ is the same (see Fig.~\ref{twoplanes}). Hence, the probability of the initial configuration resolving into the configuration shown in Fig.~\ref{pert_square23}(b) or (d) is $ (\pi - \theta)/2\pi $ for each.  By the same reasoning, the probability of the initial configuration resolving into the configuration shown in Fig.~\ref{pert_square23}(c) is $\theta/2\pi \approx 29.02\%$. 

\subsection{Square Lattice}
\label{Square Lattice}

\begin{figure}[b]
\begin{center}
\begin{tabular}{cc}
\fbox{\includegraphics[height=0.27\linewidth]{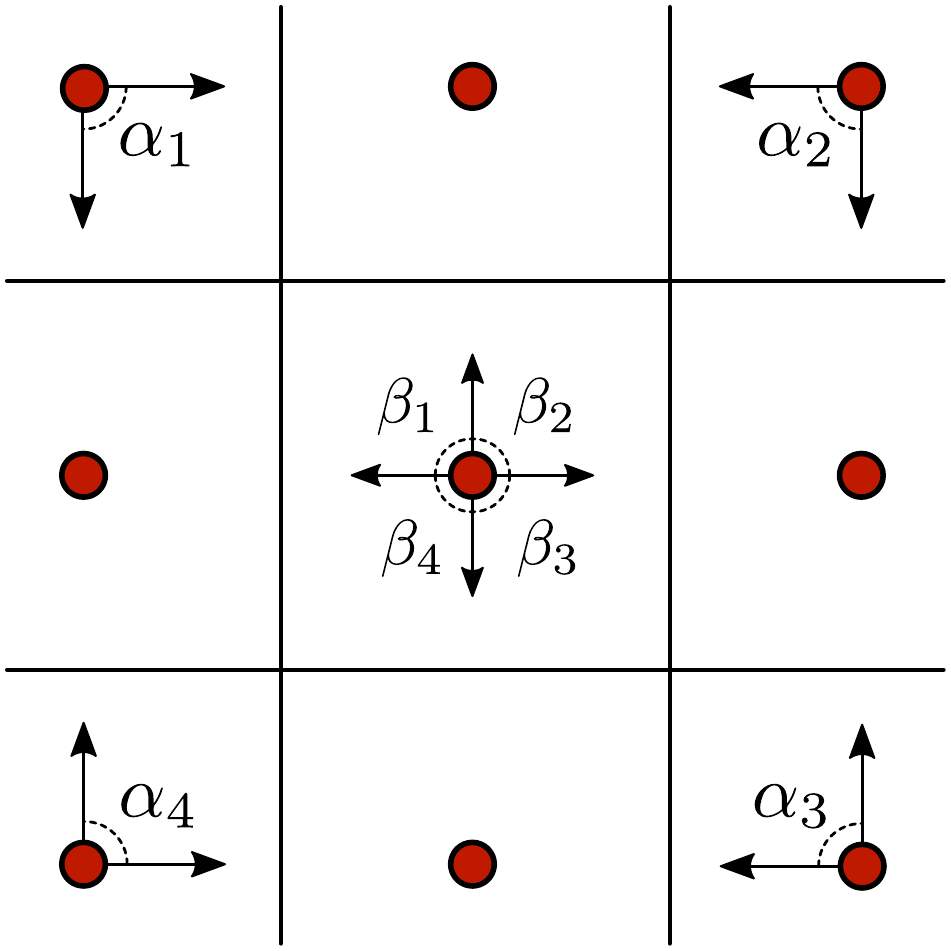}} \quad &
\fbox{\includegraphics[height=0.27\linewidth]{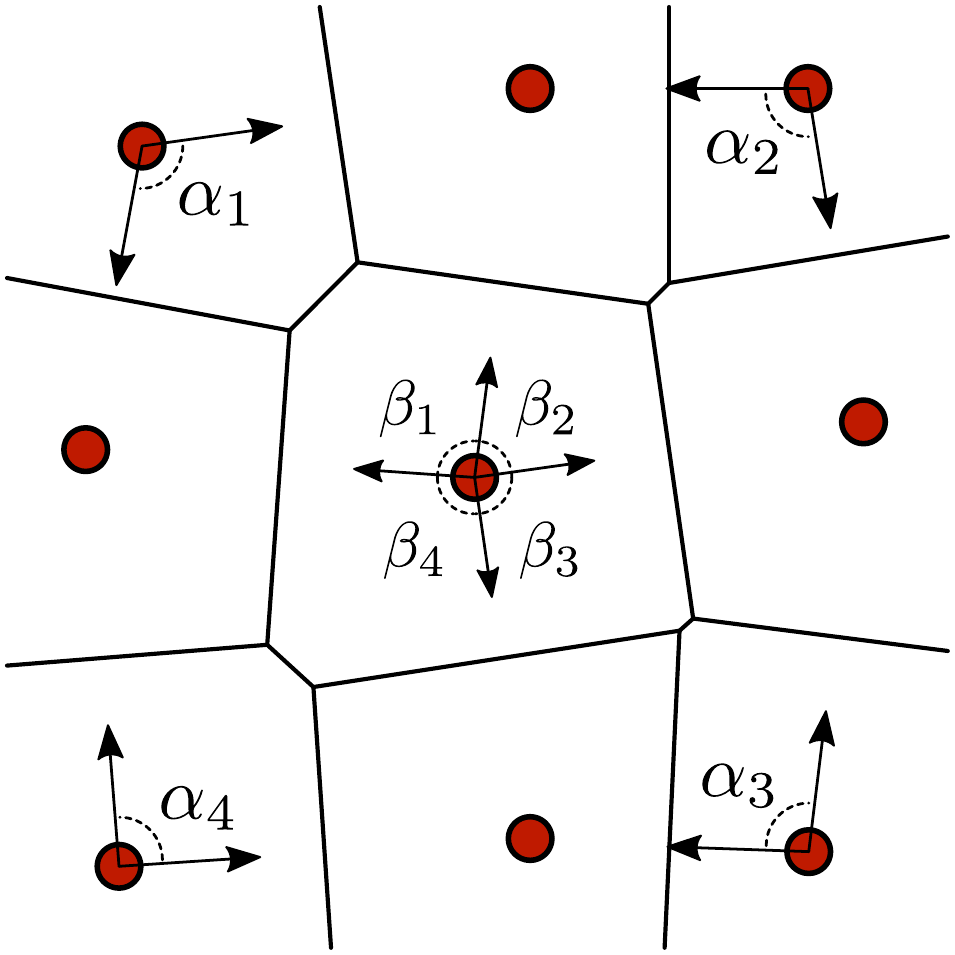}} \vspace{1mm}\\ 
(a) & (b) 
\end{tabular}
\end{center}
\caption{(a) An atom and its eight nearest neighbors in a square lattice; (b) this atom under perturbation, where the resolution of the Voronoi tessellation is determined by the sums $\alpha_i + \beta_i$ formed by neighboring atoms, as illustrated here and described in the text.\label{pert_square9}}
\end{figure}

The above approach allows us to compute $p(n)$ -- the probability that the Voronoi cell of a randomly chosen atom has $n$ edges -- in a perturbed square lattice.  Figure \ref{pert_square9} illustrates a central atom in a square lattice and its eight nearest neighbors.
The resolution of all four unstable vertices in the Voronoi tessellation can be determined by considering how the  sums of angles $\phi_i = \alpha_i + \beta_i$ change under perturbation. Following the approach above, the resolution of the entire configuration under a perturbation $\xi$ is determined by the signs of $\nabla \phi_i \cdot \xi$.

The resolution of each unstable vertex can be described by a pair of half-spaces in $\mathbb{R}^{2n}$ with $ n =  9$; in one half space $\nabla \phi_i \cdot \xi> 0$, and in the other $ \nabla\phi_i \cdot \xi < 0$.  The way in which all four unstable vertices resolve is then described by the intersection of four of these half-spaces -- one for each pair.  Under a random perturbation $\xi$, the Voronoi cell of a central atom gains one edge for each $i$ for which $\nabla\phi_i \cdot \xi>0$. The probability that the Voronoi cell of an atom in a square lattice resolves with 8 edges is the fractional size of the region in $\mathbb{R}^{18}$ in which $\nabla\phi_i \cdot \xi>0$ for all $i$.  The probability that a Voronoi cell resolves with 4 edges is equal to the fractional size of the region in $\mathbb{R}^{18}$ in which $\nabla\phi_i \cdot \xi<0$ for all $i$.

If the initial, unperturbed configuration of the atoms is
\begin{equation}
\;\; X = \Bigg( \colvecr{ -1 }{  1 } , \colvec{ 0 }{  1} , \colvec{ 1 }{  1} , \colvecr{ -1 }{  0} , \colvec{ 0 }{  0} , \colvec{ 1 }{  0} , \colvec{ -1 }{  -1} , \colvec{ 0 }{  -1} , \colvecr{ 1 }{  -1 }\Bigg), 
\end{equation}
the gradients $\nabla\phi_i$ are
\begin{align}
\nabla \phi_1 & =  \Bigg( \colvecr{1}{ -1} , \colvec{ 1}{ 1} , \colvec{ 0}{ 0} , \colvec{ -1}{ -1} , \colvecr{ 1}{ -1} , \colvec{ 0}{ 0} , \colvec{ 0}{ 0} , \colvec{ 0}{ 0} , \colvec{ 0}{ 0} \Bigg)
\\[1em]
\nabla \phi_2 & = \Bigg( \colvec{0}{ 0} , \colvecr{ -1}{ 1} , \colvec{ -1}{ -1} , \colvec{ 0}{ 0} , \colvec{ 1}{ 1} , \colvecr{ 1}{ -1} , \colvec{ 0}{ 0} , \colvec{ 0}{ 0} , \colvec{ 0}{ 0 } \Bigg) \\[1em]
\nabla \phi_3 & = \Bigg( \colvec{0}{ 0} , \colvec{ 0}{ 0} , \colvec{ 0}{ 0} , \colvec{ 0}{ 0} , \colvecr{ 1}{ -1} , \colvec{ 1}{ 1} , \colvec{ 0}{ 0} , \colvec{ -1}{ -1} , \colvecr{ 1}{ -1 } \Bigg) \\[1em]
\nabla \phi_4 & = \Bigg( \colvec{0}{ 0} , \colvec{ 0}{ 0} , \colvec{ 0}{ 0} , \colvecr{ -1}{ 1} , \colvec{-1}{ -1} , \colvec{ 0}{ 0} , \colvec{ 1}{ 1} , \colvecr{ 1}{ -1} , \colvec{ 0}{ 0 } \Bigg).
\end{align}

By consideration of their dot products, it is straightforward to see that $\nabla\phi_1$ is orthogonal to $\nabla\phi_2$ and $\nabla\phi_4$, though not to $\nabla\phi_3$.  In fact, each $\nabla\phi_i$ is orthogonal to the $\nabla\phi_i$ associated with its two nearest neighbors. For any two orthogonal vectors ${\bf w_1}, {\bf w_2} $ and a random perturbation $\xi $ chosen from a radially symmetric distribution, $P[{\bf w_1}\cdot\xi>0,{\bf w_2}\cdot\xi > 0] = P[{\bf w_1}\cdot\xi>0]P[{\bf w_2}\cdot\xi>0] $. This follows from the fact that a radially symmetric distribution is invariant under rotation.  Therefore, we can compute the joint probability $P[\nabla\phi_1 \cdot \xi>0, \nabla\phi_3 \cdot \xi>0]$ independently of the joint probability $P[\nabla\phi_2 \cdot \xi>0, \nabla\phi_4 \cdot \xi>0]$, and so the probability of a Voronoi cell resolving with 8 edges is given by:
\begin{eqnarray}
p(8) &=& P[\nabla \phi_i \cdot \xi > 0] \nonumber\\
& = & P[\nabla \phi_1 \cdot \xi > 0, \nabla \phi_3 \cdot \xi > 0] \cdot P[\nabla \phi_2 \cdot \xi > 0, \nabla \phi_4 \cdot \xi > 0].
\end{eqnarray}
Since we have shown (Section \ref{Two Nearby Vertices in a Square Lattice}) that $P[\nabla \phi_1 \cdot \xi > 0, \nabla \phi_3 \cdot \xi > 0] = P[\nabla \phi_2 \cdot \xi > 0, \nabla \phi_4 \cdot \xi > 0] = (\pi - \theta)/2\pi$, the joint probability of $\nabla \phi_i \cdot \xi > 0$ for all $i$ is $[(\pi - \theta)/2\pi]^2 \approx 4.40\%$.  Because the probability distribution of the perturbations is radially symmetric, this is also the probability $p(4)$.  This is consistent with numerical results reported below in Section \ref{numerics}.

We determine $ p(5)$ and $ p(7) $ in a similar manner. For the Voronoi cell to resolve with five edges, one vertex must resolve to gain an edge and all remaining vertices must resolve so that no other edge is gained. This can occur in four ways, one for each vertex that can gain an edge; the sum of the probabilities of these four events is:
\begin{eqnarray}
p(5) &=& P[\nabla \phi_{1} \cdot \xi > 0, \nabla \phi_{3} \cdot \xi < 0] \cdot P[\nabla \phi_{2} \cdot \xi < 0, \nabla \phi_{4} \cdot \xi < 0]  + \nonumber \\
&& P[\nabla \phi_{1} \cdot \xi < 0, \nabla \phi_{3} \cdot \xi > 0] \cdot P[\nabla \phi_{2} \cdot \xi < 0, \nabla \phi_{4} \cdot \xi < 0] + \nonumber \\
&& P[\nabla \phi_{1} \cdot \xi < 0, \nabla \phi_{3} \cdot \xi < 0] \cdot P[\nabla \phi_{2} \cdot \xi > 0, \nabla \phi_{4} \cdot \xi < 0] + \nonumber \\
&&P[\nabla \phi_{1} \cdot \xi < 0, \nabla \phi_{3} \cdot \xi < 0] \cdot P[\nabla \phi_{2} \cdot \xi < 0, \nabla \phi_{4} \cdot \xi > 0] \; \; \\
&=&\frac{\theta(\pi-\theta)}{\pi^2} \approx 24.35 \%.
\end{eqnarray}
Noting that the probability of two opposite sides both gaining an edge is the same as both not gaining an edge, clearly $p(5) = p(7)$. We can  calculate $p(6)$ by noting that we have already determined every other case (since it must resolve so that the central atom has 4 to 8 edges); i.e., $p(6) = 1  - p(4) - p(5) - p(7) - p(8) = \frac{1}{2}-\frac{\theta}{\pi} + \frac{3}{2}\left(\frac{\theta}{\pi}\right)^2  \approx 42.49 \% $.

\subsection{Rectangular Lattice}

Our approach for the square lattice naturally leads to results for rectangular lattices.  We can describe the {\it aspect ratio} of a rectangular lattice as  $ y = \|{\bf v_1}\| /\| {\bf v_2}\|$, where ${\bf v_1}$ and ${\bf v_2}$ are basis vectors for the lattice as described in Section~\ref{Classification for Two Dimensional Lattices}. Following the conventions and definitions for the square lattice, if the initial positions of the atoms are given by the configuration
\begin{equation}
X = \left( \colvec{ -1 }{ y }, \colvec{ 0 }{ y }, \colvec{ 1 }{ y }, \colvecr{ -1 }{ 0 }, \colvec{ 0 }{ 0 }, \colvec{ 1 }{ 0 }, \colvec{ -1 }{ -y }, \colvec{ 0 }{ -y }, \colvec{ 1 }{ -y } \right) ,
\end{equation}
then the gradients $ \nabla \phi_i $ are
\begin{align}
\nabla \phi_1 &= \bigg( \colvec{ 1/y}{ -1}, \colvec{ 1/y}{ 1}, \colvec{ 0}{ 0}, \colvec{ -1/y}{ -1}, \colvec{ -1/y}{ 1}, \colvec{ 0}{ 0}, \colvec{ 0}{ 0}, \colvec{ 0}{ 0}, \colvec{ 0}{ 0 } \, \bigg) \\[1em] 
\nabla \phi_2 &= \bigg( \colvec{ 0 }{ 0 }, \colvec{ -1/y }{ 1 }, \colvec{ -1/y }{ -1}, \colvec{ 0}{ 0}, \colvec{ 1/y}{ 1}, \colvec{ 1/y }{ -1 }, \colvec{ 0 }{ 0 }, \colvec{ 0 }{ 0 }, \colvec{ 0}{ 0 }\, \bigg) \\[1em] 
\nabla \phi_3 &= \bigg( \colvec{ 0 }{ 0 }, \colvec{ 0} { 0 }, \colvec{ 0 }{ 0 }, \colvec{ -1/y }{ 1 },\colvec{ -1/y }{ -1 },\colvec{ 0 }{ 0 },\colvec{ 1/y }{ 1 },\colvec{ 1/y }{ -1 }, \colvec{ 0 }{ 0 } \,  \bigg) \\[1em] 
\nabla \phi_4 &= \bigg( \colvec{ 0}{ 0}, \colvec{ 0 }{ 0 }, \colvec{ 0}{ 0}, \colvec{ 0}{ 0}, \colvec{ 1/y}{ -1}, \colvec{ 1/y}{ 1}, \colvec{ 0}{ 0}, \colvec{ -1/y}{ -1}, \colvec{ -1/y}{ 1 } \, \bigg).
\end{align}

Under a random perturbation $\xi$, the rectangular Voronoi cell of a central atom will gain an edge for each $i$ for which $\nabla\phi_i\cdot\xi > 0$, in a manner similar to that illustrated in Fig.~\ref{pert_square}.  Determining the distribution of Voronoi topologies $p(n)$ of a perturbed rectangular lattice requires computing the fractional size of the region in which $\nabla\phi_i\cdot\xi$ have particular signs.  In the general case, this fractional volume is equivalent to the solid angle subtended by the region, divided by the surface area of the unit $n$-sphere. 

\begin{figure}
\begin{center}
\includegraphics[width=0.48\linewidth]{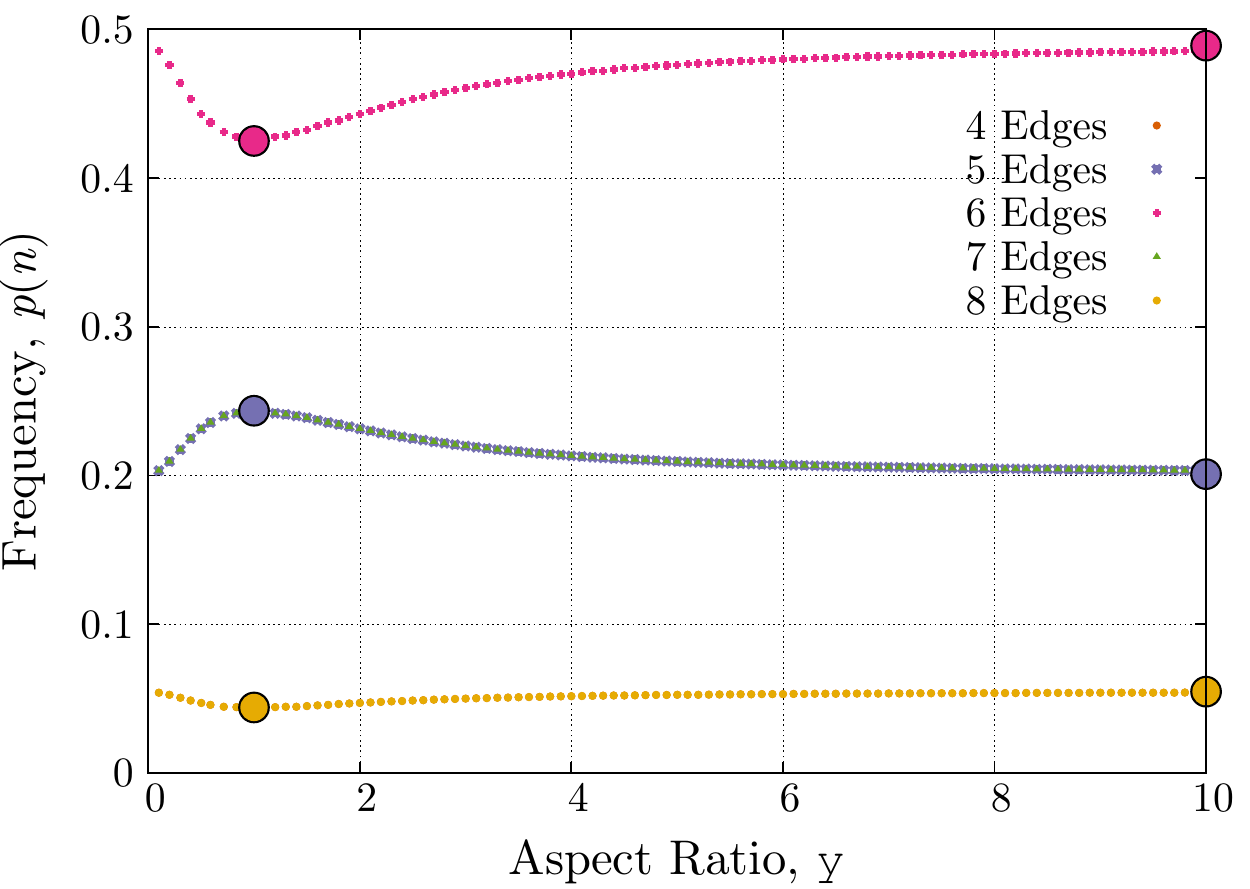}
\end{center}
\caption{Numerical integration results for an infinitesimal perturbation as a function of  rectangular lattice aspect ratio, $y$. When the aspect ratio is one, the result is the square lattice.  Large circles indicate $p(n)$ for aspect ratio 1, and also for asymptotic values as the aspect ratio approaches infinity.}
\label{recratiofunc} 
\end{figure}

No closed-form expression exists to calculate the solid angle bounded by $n$ linearly-independent vectors when $n>3$ \cite{murakami2005volume}. Nevertheless, several approaches are available to numerically compute this solid angle~\cite{ribando2006measuring}. In this paper, we compute the probabilities $ p(n) $ for each $ n $ by numerically integrating the Gaussian function over the intersection of the four half spaces. Although any radially symmetric probability function could have been used in the integration, the Gaussian was selected due to its well-understood functional form. Figure \ref{recratiofunc} shows $p(n)$ obtained from numerical integration for aspect ratios $0<y \leq 10$.  Note that $p(n) $ is the same for $y$ as for $1/y$.

It is interesting to inquire how $p(n)$ behaves  as $y$ approaches $0$ or $\infty$.  Limit analysis of the integral equations, combined with numerical integration, shows that $ p(4) = 5.46 \%$, $p(5) = 20.09\%$, $p(6) = 48.90 \%$, $p(7) = 20.09\%$, and $ p(8) = 5.46 \% $ as $y\rightarrow\infty$ or $y\rightarrow0$  (see Fig.~\ref{recratiofunc}).

\subsection{Vertex in a Honeycomb Structure}

\begin{figure}
\includegraphics[width=0.7\columnwidth]{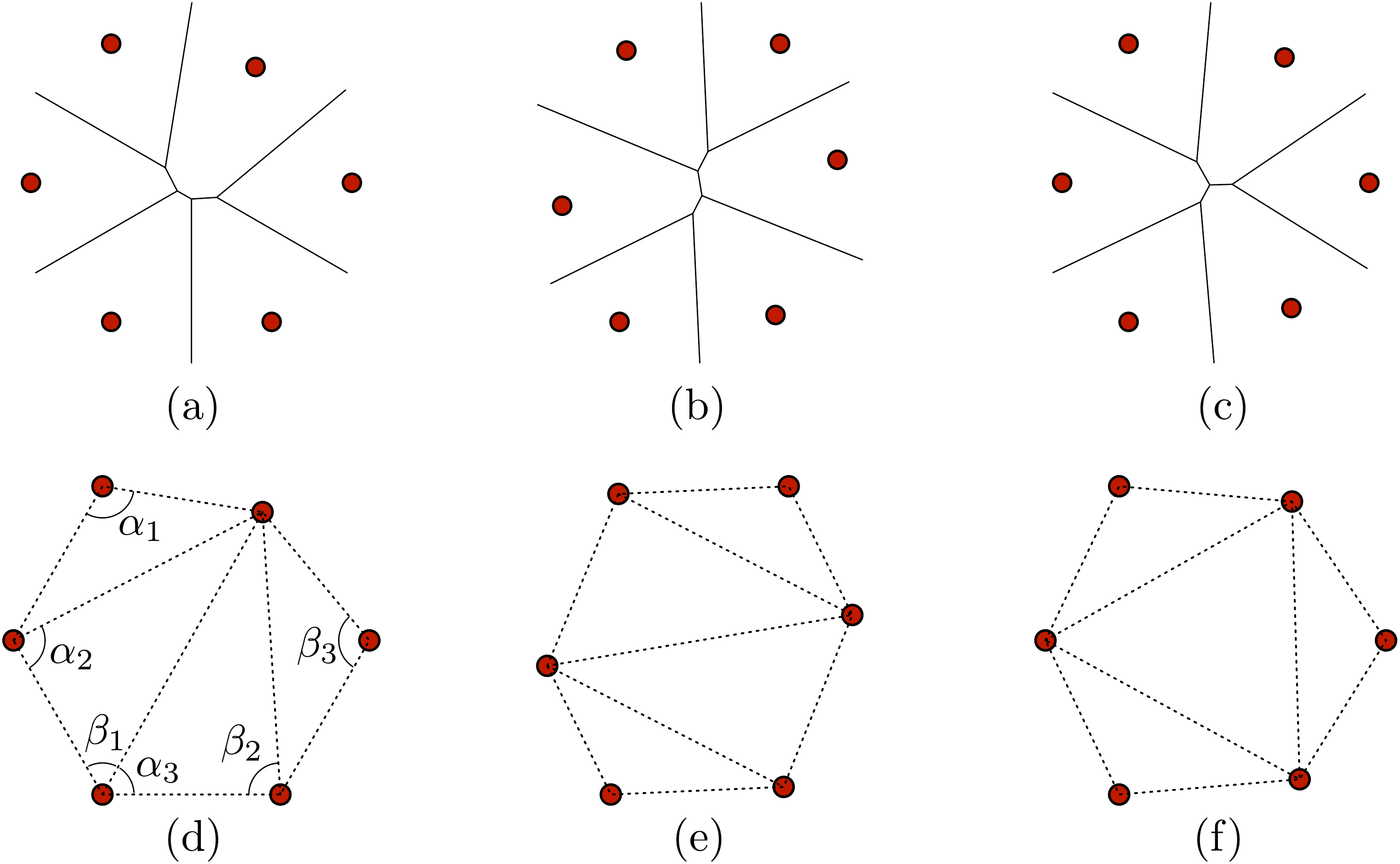}
\caption{(a-c) Resolutions of unstable vertex in the honeycomb structure; (d-f) corresponding Delaunay diagrams.
\label{pert_square6}}
\end{figure}

We next focus on the honeycomb structure shown in Fig.~\ref{bravais}(f); we note that this is not a Bravais lattice, since there are atoms not equivalent by translation.  In the center of the hexagonal arrangement of atoms is an unstable Voronoi vertex which is adjacent to six edges.  With probability one, the original unstable vertex resolves into three stable vertices, each of which is adjacent to three edges.  Such a resolution can occur in one of three ways, as illustrated in Figs.~\ref{pert_square6}(a-c).  

To compute the probabilities of each of these resolutions, we consider the initial configuration of atoms
\begin{equation}
X = \left( \colvec{ 1/2 }{ \sqrt{3}/2 }, \colvec{ -1/2 }{ -\sqrt{3}/2 }, \colvecr{ -1 }{ 0 }, \colvec{ -1/2 }{ -\sqrt{3}/2 }, \colvec{ 1/2 }{ -\sqrt{3}/2 }, \colvec{ 1 }{ 0 } \right)
\end{equation}
and the corresponding gradients $\nabla\phi_i$ 
\begin{eqnarray}
 \nabla \phi_{1} &=& \left( \colvec{ \sqrt{3}/4 }{ 3/4 }, \colvec{  0 }{  0 }, \colvec{  0 }{  0 }, \colvec{  \sqrt{3}/4 }{ 3/4 }, \colvec{  \sqrt{3}/2 }{ 3/2 }, \colvec{  -\sqrt{3} }{  0 } \right)\\
 \nabla \phi_{2} &=& \left( \colvec{ 1 }{ 2 \, \sqrt{3}  }, \colvec{  0 }{  0 }, \colvec{ 2/\sqrt{3}}{  0 }, \colvec{ -\sqrt{3}/2 }{ -3/2 }, \colvec{  -1/\sqrt{3} }{  1 }, \colvec{  0 }{  0 }\right)\\
 \nabla \phi_{3} &=& \left( \colvec{ \sqrt{3}/4 }{  3/4 }, \colvec{  \sqrt{3}/2 }{  -3/2 }, \colvec{  -\sqrt{3} }{  0 }, \colvec{  \sqrt{3}/4 }{  3/4 }, \colvec{  0 }{  0 }, \colvec{  0 }{  0 }\right),
\end{eqnarray}
where $\phi_i = \alpha_i + \beta_i$, for a resolution of the initial configuration as shown in Fig.~\ref{bravais}(f). Each of the resolutions shown in Figs.~\ref{pert_square6}(a-c) has a triplet of gradient vectors associated with it that determine the probability of the initial configuration resolving as shown. The Voronoi topology resulting from a perturbation can be determined through examination of the Delaunay diagrams.  Each topological resolution corresponds to a distinct Delaunay triangulation, the topology of which depends on satisfying a set of inequalities in terms of internal angles formed by the six neighboring atoms.  

For example, in order for the unstable vertex to resolve in the manner illustrated in Fig.~\ref{pert_square6}(a), it is necessary that $\alpha_i+\beta_i>\pi$ for all $i$; this occurs when $\nabla\phi_i\cdot\xi > 0$ for all $i$.  The probability of obtaining this resolution is equal to the fractional region of space of three intersecting half spaces, each described by $\nabla\phi_i\cdot\xi > 0$.  We can compute this fractional region by considering $\Omega$, the solid angle subtended by this region of space. For the configuration described by the Voronoi tessellation in Fig.~\ref{pert_square6}(a) and the Delaunay the in Fig.~\ref{pert_square6}(d), we compute $\Omega = 2 \, \text{tan}^{-1}\left( \sqrt{6} - 2 \right)$.  We then have:
\begin{equation}
P[\nabla \phi_i \cdot \xi > 0] = \frac{\Omega}{4 \pi} = \frac{2\,\text{tan}^{-1}\left( \sqrt{6} - 2 \right)}{4\pi} \approx 0.067,
\end{equation}
since the fractional volume of a three dimensional wedge is its solid angle divided by the surface area of a three-dimensional sphere.  Accounting for symmetries, Fig.~\ref{pert_square6}(a) can be obtained in  six equivalent ways (obtained by $60^{\circ}$ rotations), such that the probability of obtaining this topology is $6 \, P[\nabla \phi_i \cdot \xi > 0] \approx 40.34\%$.  Similar calculations show that the probability of obtaining the topology in Fig.~\ref{pert_square6}(b) is $ \frac{3}{\pi} \, \text{tan}^{-1}\left(\frac{2\,\sqrt{6}-3}{5}\right) \approx 34.66\% $ and in Fig.~\ref{pert_square6}(c) is $1/4$.

\subsection{Hexagonal, Oblique, and Rhombic Lattices}

For infinitesimal perturbations, the hexagonal, oblique and rhombic lattices are all stable, since for any empty quadrilateral, opposite angles never sum to $\pi$;  the Delaunay condition (Section \ref{Delaunay Triangulation}) shows that these are all  stable triangulations.  With probability 1, an infinitesimal perturbation cannot change this; in the limit that the magnitude of the random perturbation goes to zero, there will be a point at which the linearization $ | \nabla\phi \cdot \xi| < \gamma$ where $ \gamma $ is the difference from the original sum of opposite angles and $\pi$. Therefore, in this limit, these structures are {\it stable}. As an example, consider the hexagonal lattice from Fig.~\ref{bravais}(f), where any empty quadrilateral has a pair of opposite angles that sum to $2\pi/3$ and thus are stable for all small perturbations. 

Table \ref{summary} summarizes the analytic results from Section~\ref{Analytic Solutions}, as well as the known Poisson-Voronoi distribution.

\begin{table}
\begin{equation*}
\begin{array}{ l | c | c | c | c | c | c | c }
 & p(3) & p(4) & p(5) & p(6) & p(7) & p(8) & p(9) \\
\hline \Tstrut
\text{Square (exact)}  & 0  & [(\pi - \theta)/2\pi]^2& \frac{\theta(\pi-\theta)}{\pi^2} & \frac{1}{2}-\frac{\theta}{\pi} + \frac{3}{2}\left(\frac{\theta}{\pi}\right)^2 & \frac{\theta(\pi-\theta)}{\pi^2} & [(\pi - \theta)/2\pi]^2 &  0\\[0.25em]
\text{Square (approx.)} &   0  & {  0.0440 }& {  0.2435 } & {  0.4249 } & {  0.2435 } & {  0.0440 } &  0\\[0.25em]
\text{Rectangular (3)} &  0 & 0.0499 & 0.2198 & 0.4605 & 0.2197 & 0.0499 &  0\\[0.25em]
\text{Rectangular (10)} &  0 & 0.0539 & 0.2033 & 0.4856 & 0.2033 & 0.0539 &  0 \\[0.25em]
\text{Rectangular ($\infty$)} &  0 & 0.0546 & 0.2009 & 0.4888 & 0.2009 & 0.0546 &  0 \\[0.25em]
\text{Hexagonal} &  0 &  0 &  0 & {  1 } &  0 &  0 &  0\\[0.25em]
\text{Rhombic} &  0 &  0 &  0 & {  1 } &  0 &  0 &  0\\[0.25em]
\text{Oblique} &  0 &  0 &  0 & {  1 } &  0 &  0 &  0\\[0.25em]
\text{Poisson} & 0.0112 & 0.1068 & 0.2595 & 0.2947 & 0.1987 & 0.0897 & 0.0295 \\[0.25em]
\end{array}
\end{equation*}
\caption{Summary of results for the distribution of Voronoi topologies under infinitesimal perturbation for each Bravais lattice and for Poisson distributed points \cite{calka2003explicit}.  As explained in Section \ref{Two Nearby Vertices in a Square Lattice}, $\theta = \arccos(-1/4)$.\label{summary}}
\end{table}

\section{Simulations}
\label{numerics}

While we reported above on the effects of  infinitesimal perturbations on the topological distributions of Voronoi cells, we now examine the impact of the perturbation amplitude on the same distributions. For this purpose, we displace each atom $i$ from its initial lattice position $x_i$ by a two-dimensional independent random variable $ \Delta x_i $ chosen from a Gaussian distribution centered at the origin, and described by the probability density function
\begin{equation}
f(\Delta x_i) = \frac{1}{{2\pi \sigma^2}} e^{ -| \Delta x_i|^2 /{2\sigma^2} };
\end{equation}
the new position of each atom after the perturbation is $x_i' = x_i + \Delta x_i$.

We consider how the distribution of Voronoi topologies $ p(n) $ in the perturbed system varies with the width of the Gaussian $\sigma$. When $ \sigma$ is small, the results obtained in Section~\ref{Analytic Solutions} should be closely reproduced. When $ \sigma $ is large, $ p(n) $ approaches the Poisson-Voronoi distribution, as the distribution of atoms within any region approaches a Poisson distribution. The Poisson-Voronoi distribution is known analytically in integral form \cite{calka2003explicit} and has been evaluation numerically and through simulations \cite{brakke1986200}.

\subsection{Simulation Design}

All simulations were designed in Matlab using the standard Voronoi package and random number generator. A grid of atoms for a  lattice structure was initialized and each atom  was displaced from its lattice position by choosing random displacements in $x$ and $y$ according to a Gaussian distribution with mean $\mu = 0$ and standard deviation $\sigma$.  The standard deviation $\sigma$ was measured in units of the square root of the unit cell area: $\sqrt{\|  {\bf v_1} \times {\bf v_2} \| } $.

Each simulation considered a system of at least one million atoms; periodic boundary conditions were used to minimize finite-size effects. The simulations recorded the fraction of Voronoi cells with a particular topology. 

\subsection{Square Lattice}
\begin{figure}
\centering
\includegraphics[width=0.48\linewidth]{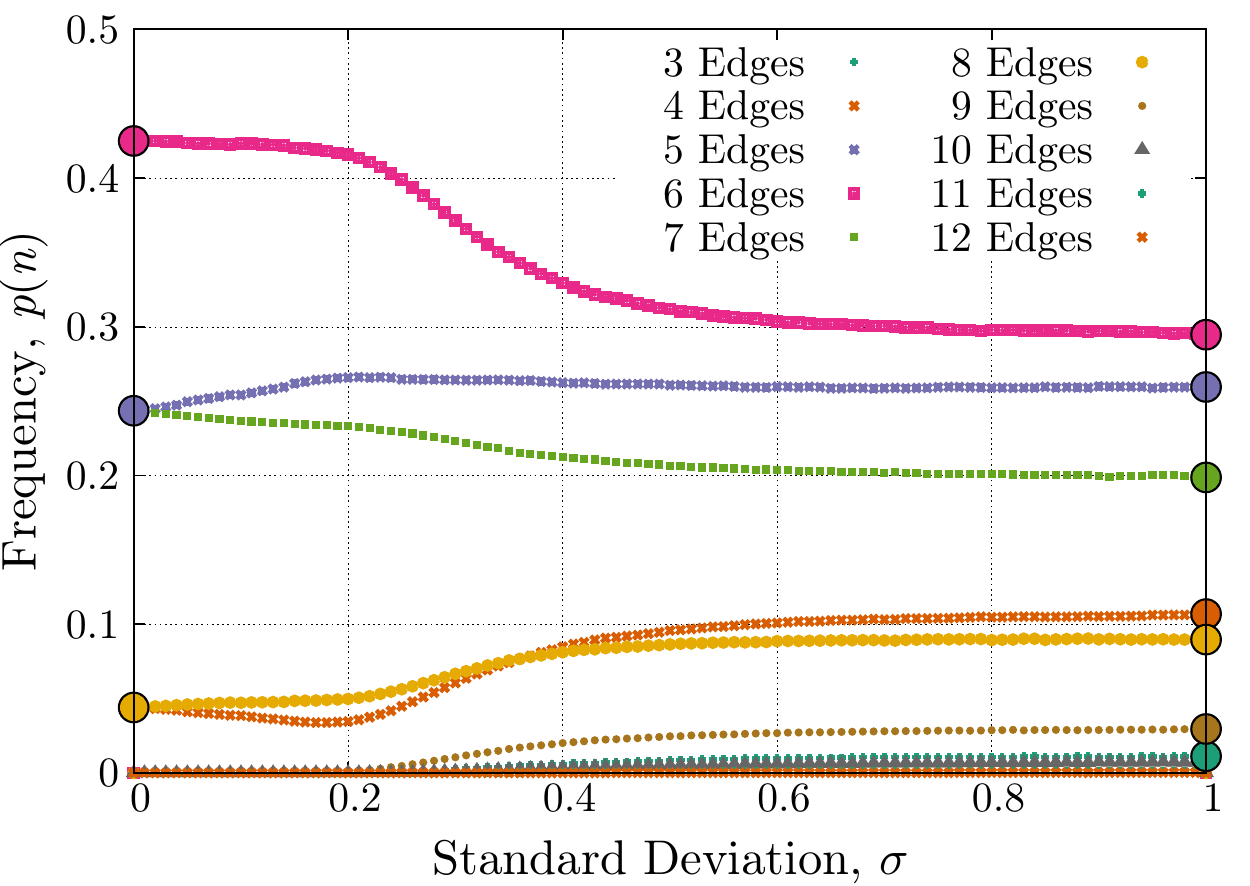}
\caption{Frequencies of $n$-sided Voronoi cells for $0 < \sigma \leq 1 $ for the square lattice; large circles indicate asymptotic values.}
\label{SqLattice}
\end{figure}

Figure \ref{SqLattice} shows the distribution of Voronoi topologies $p(n)$ of a perturbed square lattice for $0 < \sigma \leq 1 $, obtained through numerical simulation as described above.  Simulation data for small $\sigma$ are consistent with the analytical results reported in Section \ref{Square Lattice}.  The symmetry of the analytic result is quickly broken as $ \sigma $ increases.  Whereas $p(5)=p(7)$ and $p(4)=p(8)$ in the limit as $\sigma$ approaches 0, for finite $\sigma$, this is not the case.  Moreover, where $p(n)=0$ for $n=3$ and $n>8$ in the small $\sigma$ limit, these numbers become positive for finite $\sigma$.  As $ \sigma $ increases, $ p(n) $ evolves monotonically toward the Poisson-Voronoi distribution except for $ p(4) $ which has an inflection point near $ \sigma = 0.3$. The fraction of Voronoi cells with $ n>8 $ edges increases rapidly between $ \sigma=0.2 $ and $ \sigma=0.4$, and then levels off by $ \sigma=0.5$. By $\sigma = 1$, it is difficult to distinguish the distribution $p(n)$ from the Poisson-Voronoi distribution.

\subsection{Rectangular Lattice}

\begin{figure}
\begin{center}
\begin{tabular}{cc}
\includegraphics[width=0.48\linewidth]{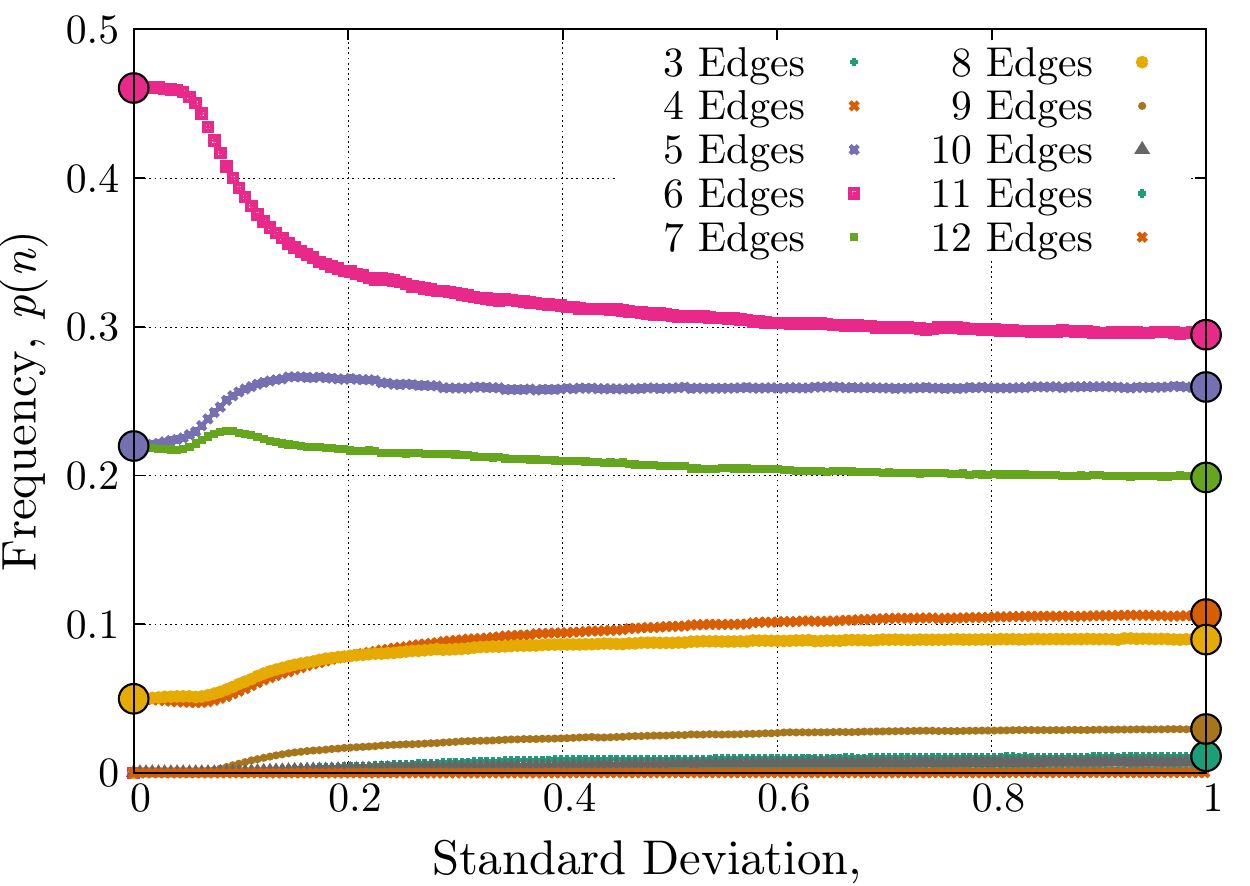} &
\includegraphics[width=0.48\linewidth]{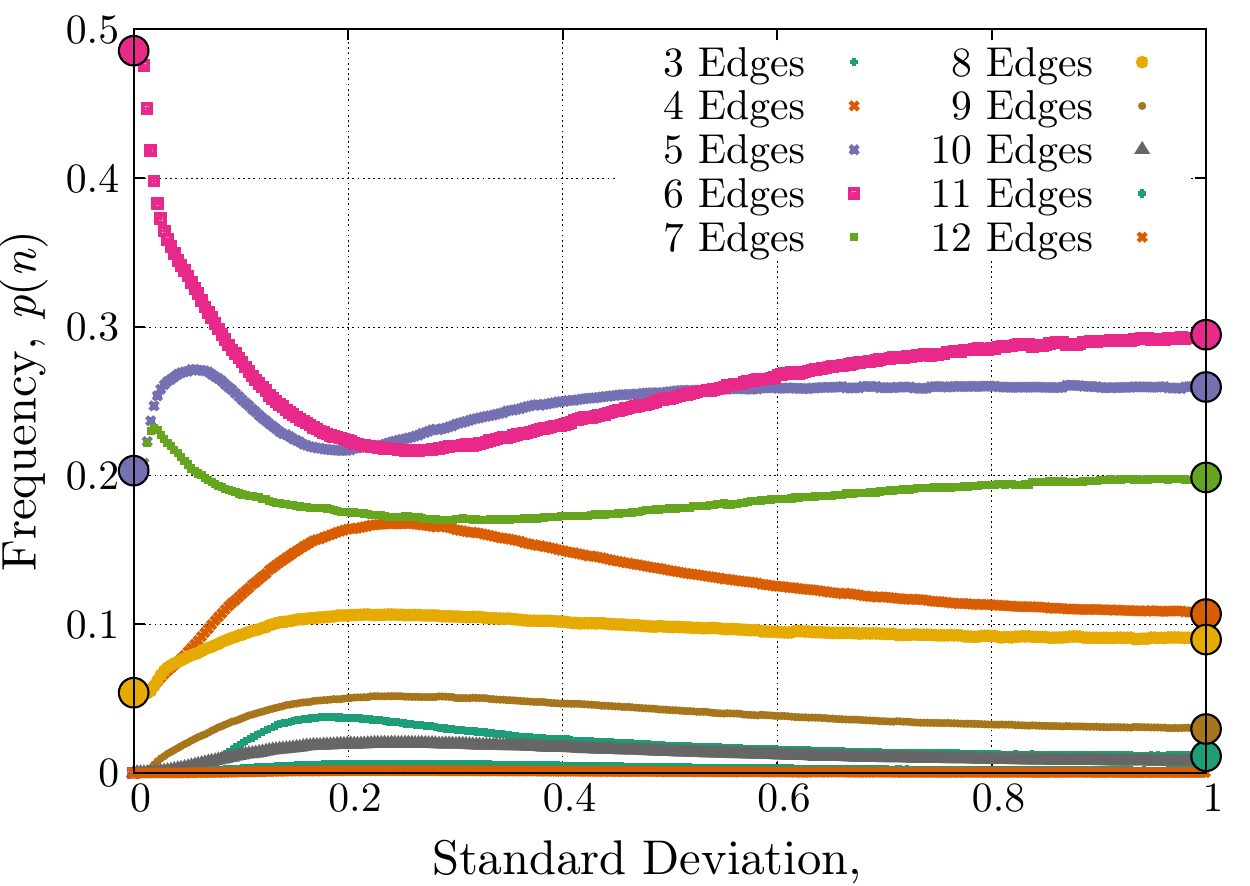} \vspace{1mm}\\ 
(a) Aspect ratio 3 & (b) Aspect ratio 10
\end{tabular}
\end{center}
\caption{Frequencies of $n$-sided Voronoi cells for $0 < \sigma \leq 1 $ for rectangular lattices of aspect ratio 3 and 10 respectively. \label{RecLattice}}
\end{figure}

Figure \ref{RecLattice} shows the distribution of Voronoi topologies $p(n)$ of rectangular lattices of aspect ratio three and ten respectively for $0 < \sigma \leq 1 $. For low $ \sigma $, the numerical results are consistent with the analytic results obtained for infinitesimal perturbations. Qualitatively, the Voronoi cells with 4, 6, and 8 edges are favored more heavily than they are in the square lattice.  Voronoi cells with 5 and 7 edges become more frequent until an abrupt change around $ \sigma = 0.2 $ for aspect ratio three and $ \sigma = 0.1 $ for aspect ratio ten. While $p(n)$ is monotonic for most $ n $ in the square lattice, $p(n)$ is not monotonic for most $n$ in the rectangle lattice at  aspect ratio ten. As the aspect ratio increases, the rectangle lattices show more significant inflections between the low and high aspect ratio limits. For example, there is a sudden peak of three-sided Voronoi cells near $\sigma = 0.5$ that disappears at higher $\sigma$. The convergence to the Poisson-Voronoi distribution is also slower as the aspect ratio increases.

\subsection{Hexagonal Lattice}

\begin{figure}
\begin{center}
\includegraphics[width=0.48\linewidth]{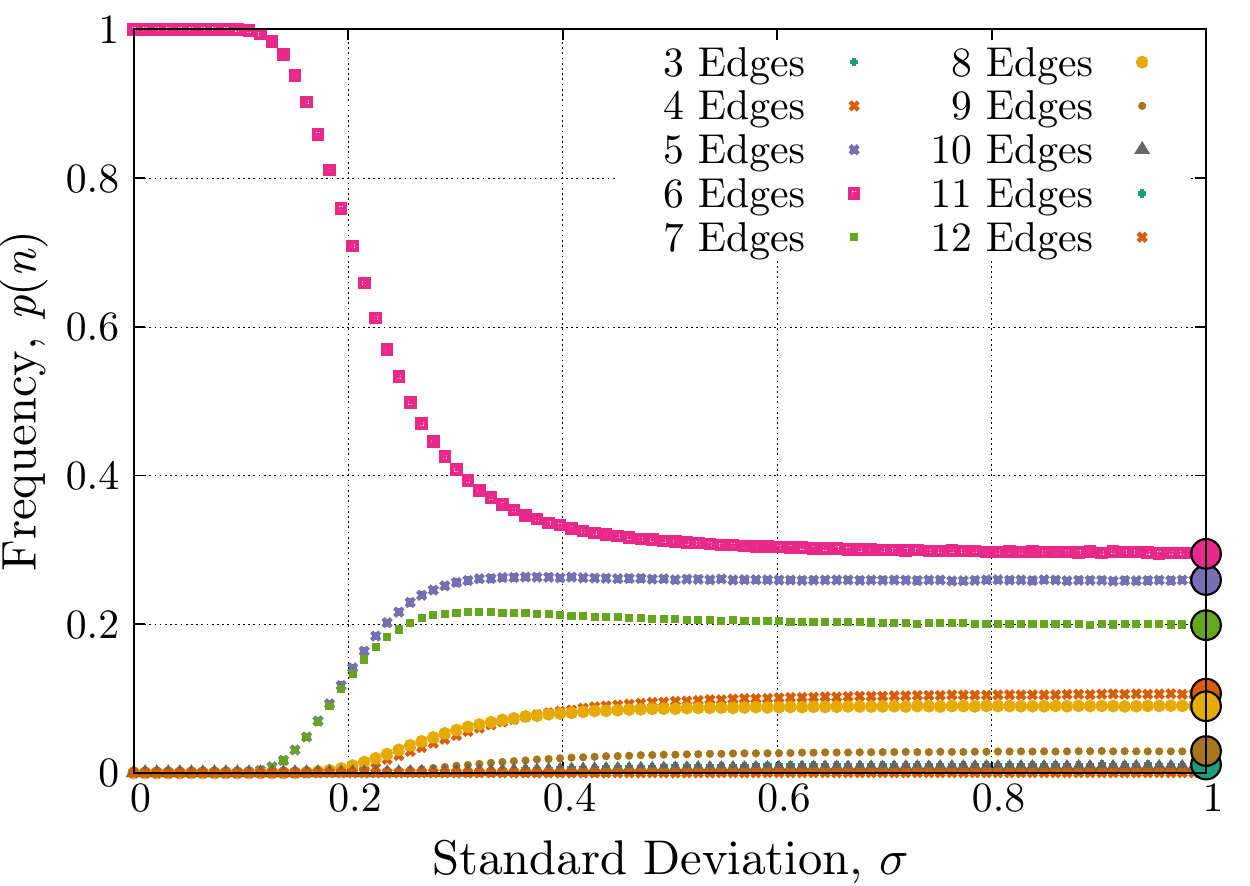} 
\end{center}
\caption{Frequencies of $n$-sided Voronoi cells for $0 < \sigma \leq 1 $ for the hexagonal lattice; large circles indicate asymptotic values.}
\label{TriLattice}
\end{figure}

Figure~\ref{TriLattice} shows the distribution of Voronoi topologies $p(n)$ for the hexagonal lattice, obtained through numerical simulation, as a function of $\sigma$.  The distribution $p(n)$ is relatively insensitive to changes of $\sigma$ at both low and high $\sigma$ limits. The switch between these limits is abrupt; it occurs largely within the limits $0.1<\sigma<0.3$.  As a result, the distribution $p(n)$ varies monotonically for most $n$ between the small and large $\sigma$ limits.

\subsection{Honeycomb Structure}

\begin{figure}
\centering
\includegraphics[width=0.48\linewidth]{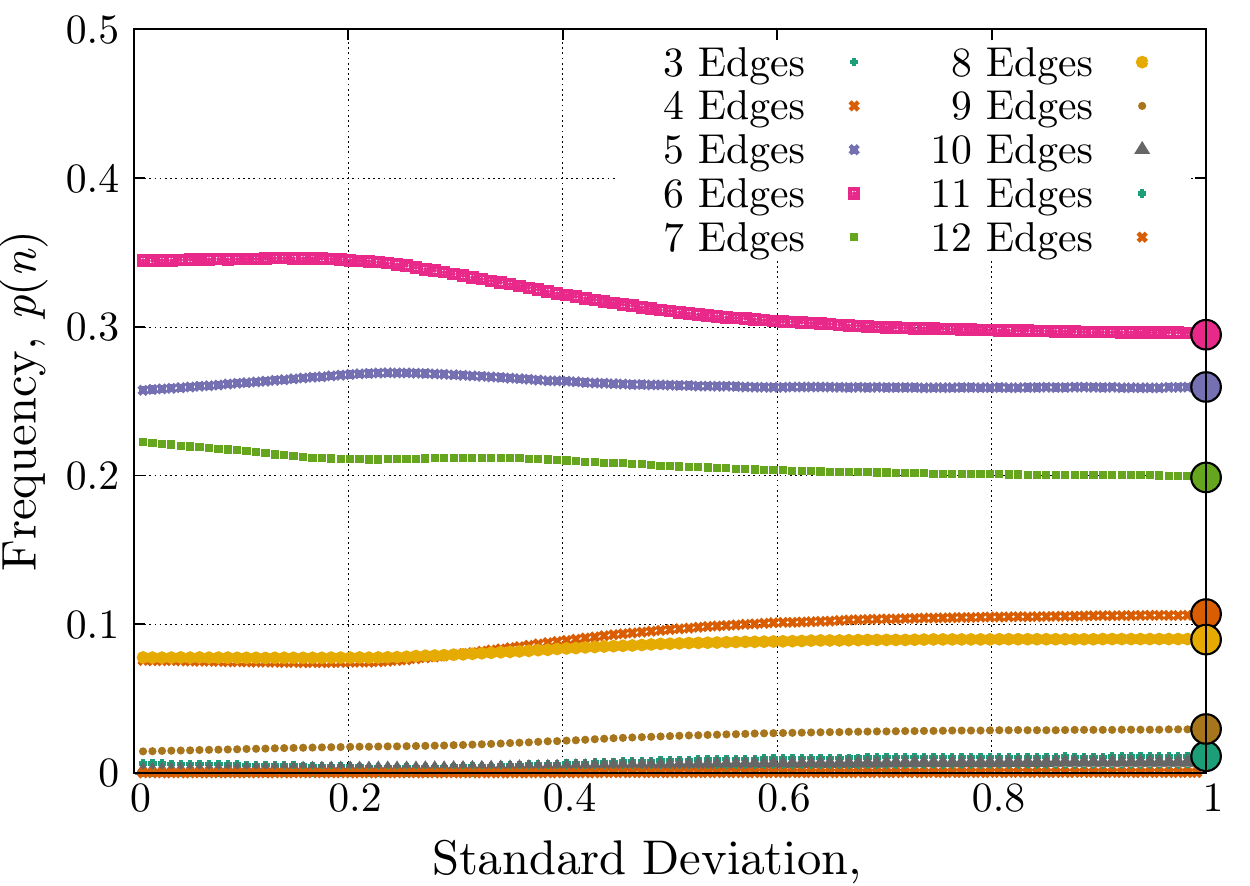}
\caption{Frequencies of $n$-sided Voronoi cells for $0 < \sigma \leq 1 $ for the honeycomb structure; large circles indicate asymptotic values.}
\label{HoneycombLattice}
\end{figure}

Figure \ref{HoneycombLattice} shows the distribution of Voronoi topologies of the honeycomb structure for $0 < \sigma \leq 1 $ as obtained through numerical simulation.  The variation of $p(n)$ with $\sigma$ is ``gentle'' relative to the lattices discussed above.  This might result from the similarity between the low- and high-$\sigma$ frequencies $p(n)$.  In the honeycomb structure, even as $\sigma $ approaches 0 there is a large diversity of topological types, with non-zero frequencies for $n=3$ and $9 \leq n \leq 12$; in contrast, for other lattices these frequencies are all zero as $\sigma $ approaches 0. 

\section{Conclusions}

In this  paper, we explored how perturbations affect the distribution of Voronoi cell topologies for the two-dimensional Bravais lattices and the honeycomb structure. Such results are important for describing crystalline materials at finite temperature. Of particular interest are those structures for which  the zero-temperature Voronoi topology distributions are unstable with respect to infinitesimal perturbations. In applications to crystalline materials, it is the small-perturbation limit that is of most interest at temperatures well below the melting point.  Even approaching the melting temperature, $\sigma$ rarely exceeds $0.05-0.15$ \cite{lawson2009physics}.

Voronoi cell descriptions of the local  structure  have been growing in importance for material characterization, both for atomic-scale simulations and in experiment.  Although here we focus on two-dimensional lattices, the same methods can be applied in three dimensions, where most materials and structures live.  Indeed a similar approach has been used in recent work \cite{topframework} to characterize disorder in three-dimensional atomic systems.

While other papers have focused on topological distributions for Poisson distributed points, our paper considers the topology in systems which have much more order, but are not strictly periodic.  The most significant achievement in this paper is the summary of exact results for the distribution of Voronoi topologies of all two-dimensional Bravais lattices under random infinitesimal perturbations chosen from any radially symmetric distribution (see Table \ref{summary}). Because we present numerical data for finite perturbations, these results also show how the topology transitions from that of a perfect lattice to a perfectly disordered (Poisson) system.  While the present study focuses on uncorrelated perturbations in the Einstein-sense, it would be of interest to see how accurate this description is for realistic atomic systems in which vibrations are correlated, as described by the phonon spectra.

\bibliographystyle{apsrev4-1.bst}
\bibliography{refs}       

\begin{thebibliography}{17}%
\makeatletter
\providecommand \@ifxundefined [1]{%
 \@ifx{#1\undefined}
}%
\providecommand \@ifnum [1]{%
 \ifnum #1\expandafter \@firstoftwo
 \else \expandafter \@secondoftwo
 \fi
}%
\providecommand \@ifx [1]{%
 \ifx #1\expandafter \@firstoftwo
 \else \expandafter \@secondoftwo
 \fi
}%
\providecommand \natexlab [1]{#1}%
\providecommand \enquote  [1]{``#1''}%
\providecommand \bibnamefont  [1]{#1}%
\providecommand \bibfnamefont [1]{#1}%
\providecommand \citenamefont [1]{#1}%
\providecommand \href@noop [0]{\@secondoftwo}%
\providecommand \href [0]{\begingroup \@sanitize@url \@href}%
\providecommand \@href[1]{\@@startlink{#1}\@@href}%
\providecommand \@@href[1]{\endgroup#1\@@endlink}%
\providecommand \@sanitize@url [0]{\catcode `\\12\catcode `\$12\catcode
  `\&12\catcode `\#12\catcode `\^12\catcode `\_12\catcode `\%12\relax}%
\providecommand \@@startlink[1]{}%
\providecommand \@@endlink[0]{}%
\providecommand \url  [0]{\begingroup\@sanitize@url \@url }%
\providecommand \@url [1]{\endgroup\@href {#1}{\urlprefix }}%
\providecommand \urlprefix  [0]{URL }%
\providecommand \Eprint [0]{\href }%
\providecommand \doibase [0]{http://dx.doi.org/}%
\providecommand \selectlanguage [0]{\@gobble}%
\providecommand \bibinfo  [0]{\@secondoftwo}%
\providecommand \bibfield  [0]{\@secondoftwo}%
\providecommand \translation [1]{[#1]}%
\providecommand \BibitemOpen [0]{}%
\providecommand \bibitemStop [0]{}%
\providecommand \bibitemNoStop [0]{.\EOS\space}%
\providecommand \EOS [0]{\spacefactor3000\relax}%
\providecommand \BibitemShut  [1]{\csname bibitem#1\endcsname}%
\let\auto@bib@innerbib\@empty
\bibitem [{\citenamefont {Rivier}(1983)}]{rivier1983statistical}%
  \BibitemOpen
  \bibfield  {author} {\bibinfo {author} {\bibfnamefont {N.}~\bibnamefont
  {Rivier}},\ }\href@noop {} {\bibfield  {journal} {\bibinfo  {journal} {Helv.
  Phys. Acta}\ }\textbf {\bibinfo {volume} {56}},\ \bibinfo {pages} {307}
  (\bibinfo {year} {1983})}\BibitemShut {NoStop}%
\bibitem [{\citenamefont {Orlandini}\ and\ \citenamefont
  {Whittington}(2007)}]{orlandini2007statistical}%
  \BibitemOpen
  \bibfield  {author} {\bibinfo {author} {\bibfnamefont {E.}~\bibnamefont
  {Orlandini}}\ and\ \bibinfo {author} {\bibfnamefont {S.~G.}\ \bibnamefont
  {Whittington}},\ }\href@noop {} {\bibfield  {journal} {\bibinfo  {journal}
  {Rev. Mod. Phys.}\ }\textbf {\bibinfo {volume} {79}},\ \bibinfo {pages} {611}
  (\bibinfo {year} {2007})}\BibitemShut {NoStop}%
\bibitem [{\citenamefont {Seong}\ \emph {et~al.}(2012)\citenamefont {Seong},
  \citenamefont {Salafia},\ and\ \citenamefont
  {Vvedensky}}]{seong2012statistical}%
  \BibitemOpen
  \bibfield  {author} {\bibinfo {author} {\bibfnamefont {R.-K.}\ \bibnamefont
  {Seong}}, \bibinfo {author} {\bibfnamefont {C.~M.}\ \bibnamefont {Salafia}},
  \ and\ \bibinfo {author} {\bibfnamefont {D.~D.}\ \bibnamefont {Vvedensky}},\
  }\href@noop {} {\bibfield  {journal} {\bibinfo  {journal} {Phil. Mag.}\
  }\textbf {\bibinfo {volume} {92}},\ \bibinfo {pages} {230} (\bibinfo {year}
  {2012})}\BibitemShut {NoStop}%
\bibitem [{\citenamefont {Mason}\ \emph {et~al.}(2012)\citenamefont {Mason},
  \citenamefont {Lazar}, \citenamefont {MacPherson},\ and\ \citenamefont
  {Srolovitz}}]{mason2012statistical}%
  \BibitemOpen
  \bibfield  {author} {\bibinfo {author} {\bibfnamefont {J.~K.}\ \bibnamefont
  {Mason}}, \bibinfo {author} {\bibfnamefont {E.~A.}\ \bibnamefont {Lazar}},
  \bibinfo {author} {\bibfnamefont {R.~D.}\ \bibnamefont {MacPherson}}, \ and\
  \bibinfo {author} {\bibfnamefont {D.~J.}\ \bibnamefont {Srolovitz}},\
  }\href@noop {} {\bibfield  {journal} {\bibinfo  {journal} {Phys. Rev. E}\
  }\textbf {\bibinfo {volume} {86}},\ \bibinfo {pages} {051128} (\bibinfo
  {year} {2012})}\BibitemShut {NoStop}%
\bibitem [{\citenamefont {Lazar}\ \emph {et~al.}(2013)\citenamefont {Lazar},
  \citenamefont {Mason}, \citenamefont {MacPherson},\ and\ \citenamefont
  {Srolovitz}}]{2013lazar}%
  \BibitemOpen
  \bibfield  {author} {\bibinfo {author} {\bibfnamefont {E.~A.}\ \bibnamefont
  {Lazar}}, \bibinfo {author} {\bibfnamefont {J.~K.}\ \bibnamefont {Mason}},
  \bibinfo {author} {\bibfnamefont {R.~D.}\ \bibnamefont {MacPherson}}, \ and\
  \bibinfo {author} {\bibfnamefont {D.~J.}\ \bibnamefont {Srolovitz}},\
  }\href@noop {} {\bibfield  {journal} {\bibinfo  {journal} {Phys. Rev. E}\
  }\textbf {\bibinfo {volume} {88}},\ \bibinfo {pages} {063309} (\bibinfo
  {year} {2013})}\BibitemShut {NoStop}%
\bibitem [{\citenamefont {Holroyd}\ and\ \citenamefont
  {Soo}(2013)}]{holroyd2013insertion}%
  \BibitemOpen
  \bibfield  {author} {\bibinfo {author} {\bibfnamefont {A.~E.}\ \bibnamefont
  {Holroyd}}\ and\ \bibinfo {author} {\bibfnamefont {T.}~\bibnamefont {Soo}},\
  }\href@noop {} {\bibfield  {journal} {\bibinfo  {journal} {Electron. J.
  Probab.}\ }\textbf {\bibinfo {volume} {18}},\ \bibinfo {pages} {1} (\bibinfo
  {year} {2013})}\BibitemShut {NoStop}%
\bibitem [{\citenamefont {Peres}\ and\ \citenamefont
  {Sly}(2014)}]{peres2014rigidity}%
  \BibitemOpen
  \bibfield  {author} {\bibinfo {author} {\bibfnamefont {Y.}~\bibnamefont
  {Peres}}\ and\ \bibinfo {author} {\bibfnamefont {A.}~\bibnamefont {Sly}},\
  }\href@noop {} {\bibfield  {journal} {\bibinfo  {journal} {arXiv preprint
  arXiv:1409.4490}\ } (\bibinfo {year} {2014})}\BibitemShut {NoStop}%
\bibitem [{\citenamefont {Kittel}\ and\ \citenamefont
  {McEuen}(1976)}]{kittel1976introduction}%
  \BibitemOpen
  \bibfield  {author} {\bibinfo {author} {\bibfnamefont {C.}~\bibnamefont
  {Kittel}}\ and\ \bibinfo {author} {\bibfnamefont {P.}~\bibnamefont
  {McEuen}},\ }\href@noop {} {\emph {\bibinfo {title} {Introduction to Solid
  State Physics}}},\ Vol.~\bibinfo {volume} {8}\ (\bibinfo  {publisher} {Wiley
  New York},\ \bibinfo {year} {1976})\BibitemShut {NoStop}%
\bibitem [{\citenamefont {West}\ \emph {et~al.}(2001)\citenamefont {West} \emph
  {et~al.}}]{west2001introduction}%
  \BibitemOpen
  \bibfield  {author} {\bibinfo {author} {\bibfnamefont {D.~B.}\ \bibnamefont
  {West}} \emph {et~al.},\ }\href@noop {} {\emph {\bibinfo {title}
  {Introduction to Graph Theory}}},\ Vol.~\bibinfo {volume} {2}\ (\bibinfo
  {publisher} {Prentice Hall Upper Saddle River},\ \bibinfo {year}
  {2001})\BibitemShut {NoStop}%
\bibitem [{\citenamefont {Bravais}\ and\ \citenamefont
  {de~Beaumont}(1866)}]{bravais1866etudes}%
  \BibitemOpen
  \bibfield  {author} {\bibinfo {author} {\bibfnamefont {A.}~\bibnamefont
  {Bravais}}\ and\ \bibinfo {author} {\bibfnamefont {L.~{\'E}.}\ \bibnamefont
  {de~Beaumont}},\ }\href@noop {} {\emph {\bibinfo {title} {Etudes
  cristallographiques: M{\'e}moire sur les syst{\`e}mes form{\'e}s par des
  points distribu{\'e}s r{\'e}guli{\`e}rement sur un plan ou dans l'espace}}}\
  (\bibinfo  {publisher} {Gauthiers-Villars},\ \bibinfo {year}
  {1866})\BibitemShut {NoStop}%
\bibitem [{\citenamefont {De~Berg}\ \emph {et~al.}(2000)\citenamefont
  {De~Berg}, \citenamefont {Van~Kreveld}, \citenamefont {Overmars},\ and\
  \citenamefont {Schwarzkopf}}]{de2000computational}%
  \BibitemOpen
  \bibfield  {author} {\bibinfo {author} {\bibfnamefont {M.}~\bibnamefont
  {De~Berg}}, \bibinfo {author} {\bibfnamefont {M.}~\bibnamefont
  {Van~Kreveld}}, \bibinfo {author} {\bibfnamefont {M.}~\bibnamefont
  {Overmars}}, \ and\ \bibinfo {author} {\bibfnamefont {O.~C.}\ \bibnamefont
  {Schwarzkopf}},\ }\href@noop {} {\emph {\bibinfo {title} {Computational
  Geometry}}}\ (\bibinfo  {publisher} {Springer},\ \bibinfo {year}
  {2000})\BibitemShut {NoStop}%
\bibitem [{\citenamefont {Murakami}\ and\ \citenamefont
  {Yano}(2005)}]{murakami2005volume}%
  \BibitemOpen
  \bibfield  {author} {\bibinfo {author} {\bibfnamefont {J.}~\bibnamefont
  {Murakami}}\ and\ \bibinfo {author} {\bibfnamefont {M.}~\bibnamefont
  {Yano}},\ }\href@noop {} {\bibfield  {journal} {\bibinfo  {journal} {Comm.
  Anal. Geom.}\ }\textbf {\bibinfo {volume} {13}},\ \bibinfo {pages} {379}
  (\bibinfo {year} {2005})}\BibitemShut {NoStop}%
\bibitem [{\citenamefont {Ribando}(2006)}]{ribando2006measuring}%
  \BibitemOpen
  \bibfield  {author} {\bibinfo {author} {\bibfnamefont {J.~M.}\ \bibnamefont
  {Ribando}},\ }\href@noop {} {\bibfield  {journal} {\bibinfo  {journal}
  {Discrete Comput. Geom.}\ }\textbf {\bibinfo {volume} {36}},\ \bibinfo
  {pages} {479} (\bibinfo {year} {2006})}\BibitemShut {NoStop}%
\bibitem [{\citenamefont {Calka}(2003)}]{calka2003explicit}%
  \BibitemOpen
  \bibfield  {author} {\bibinfo {author} {\bibfnamefont {P.}~\bibnamefont
  {Calka}},\ }\href@noop {} {\bibfield  {journal} {\bibinfo  {journal} {Adv.
  Appl. Probab.}\ }\textbf {\bibinfo {volume} {35}},\ \bibinfo {pages} {863}
  (\bibinfo {year} {2003})}\BibitemShut {NoStop}%
\bibitem [{\citenamefont {Brakke}(1986)}]{brakke1986200}%
  \BibitemOpen
  \bibfield  {author} {\bibinfo {author} {\bibfnamefont {K.}~\bibnamefont
  {Brakke}},\ }\href@noop {} {\enquote {\bibinfo {title} {{200,000,000 Random
  Voronoi Polygons}},}\ } (\bibinfo {year} {1986})\BibitemShut {NoStop}%
\bibitem [{\citenamefont {Lawson}(2009)}]{lawson2009physics}%
  \BibitemOpen
  \bibfield  {author} {\bibinfo {author} {\bibfnamefont {A.~C.}\ \bibnamefont
  {Lawson}},\ }\href@noop {} {\bibfield  {journal} {\bibinfo  {journal} {Phil.
  Mag.}\ }\textbf {\bibinfo {volume} {89}},\ \bibinfo {pages} {1757} (\bibinfo
  {year} {2009})}\BibitemShut {NoStop}%
\bibitem [{\citenamefont {Lazar}\ \emph {et~al.}(2015)\citenamefont {Lazar},
  \citenamefont {Han},\ and\ \citenamefont {Srolovitz}}]{topframework}%
  \BibitemOpen
  \bibfield  {author} {\bibinfo {author} {\bibfnamefont {E.~A.}\ \bibnamefont
  {Lazar}}, \bibinfo {author} {\bibfnamefont {J.}~\bibnamefont {Han}}, \ and\
  \bibinfo {author} {\bibfnamefont {D.~J.}\ \bibnamefont {Srolovitz}},\
  }\href@noop {} {\bibfield  {journal} {\bibinfo  {journal} {arXiv preprint
  arXiv:1508.05937}\ } (\bibinfo {year} {2015})}\BibitemShut {NoStop}%
\end{thebibliography}%

\end{document}